\documentclass[preprint,showpacs,preprintnumbers,amsmath,amssymb]{revtex4}

\usepackage{graphicx}\graphicspath{./graph/}%
\usepackage{dcolumn}
\usepackage{bm}
\usepackage{color}
\usepackage{verbatim}
\usepackage{amssymb}
\usepackage{amsfonts}
\usepackage{latexsym}

\definecolor{red}{rgb}{1,0,0}
\definecolor{green}{rgb}{0,1,0}
\definecolor{blue}{rgb}{0,0,1}

\newcommand{\be}{\begin{equation}}
\newcommand{\ee}{\end{equation}}
\newcommand{\bea}{\begin{eqnarray}}
\newcommand{\eea}{\end{eqnarray}}
\newcommand{\bdm}{\begin{displaymath}}
\newcommand{\edm}{\end{displaymath}}

\newcommand\Jb{{\bf J}}
\newcommand\Rbo{{\bf R}_0}

\newcommand\Eb{{\bf E}}
\newcommand\Hb{{\bf H}}
\newcommand\ptl{\partial}
\newcommand\dpr{^{\prime\prime}}
\newcommand\pr{^\prime}

\newcommand\ra{\rightarrow}
\newcommand\Rb{{\bf R}}
\newcommand\rb{{\bf r}}
\newcommand\nb{{\bf n}}
\newcommand\tb{{\bf t}}

\newcommand\Sb{{\bf S}}

\newcommand\es{\hat E_{sp}}
\newcommand\ex{\hat E_{xp}}
\newcommand\ey{\hat E_{yp}}
\newcommand\hs{\hat H_{sp}}
\newcommand\hx{\hat H_{xp}}
\newcommand\hy{\hat H_{yp}}

\newcommand\js{\hat J_{sp}}
\newcommand\jx{\hat J_{xp}}
\newcommand\jy{\hat J_{yp}}
\newcommand\rh{\hat \rho_p}
\newcommand\fp{\hat F_p}

\newcommand\alf{\alpha_p}

\newcommand\un{u^{(n)}}

\newcommand\us{{\bf e}_s}
\newcommand\ux{{\bf e}_x}
\newcommand\uy{{\bf e}_y}
\newcommand\curl{\rm curl}

\newcommand\rep{\rm Re}
\begin{document}
\title{An Efficient Computation of Coherent Synchrotron Radiation in a Rectangular Chamber, Applied to Resistive Wall Heating}
\author{Robert L. Warnock }
\email{warnock@slac.stanford.edu}
\affiliation{SLAC National Accelerator Laboratory, Stanford University, Menlo Park, CA 94025, USA}
\author{David A. Bizzozero}
\email{bizzozero@temf.tu-darmstadt.de}
\affiliation{Department of Mathematics and Statistics, University of New Mexico, Albuquerque, NM 87131, USA}

\altaffiliation{ Inst. f. Theorie Elektromagnetischer Felder, Technische Universit\"at Darmstadt,
 Darmstadt, Germany.}

\begin{abstract}
We study coherent synchrotron radiation (CSR) in a perfectly conducting vacuum chamber of rectangular cross section, in a formalism allowing an arbitrary sequence of bends and straight sections. We apply the paraxial method in the frequency domain, with a Fourier development in the vertical coordinate but with no other mode expansions. A line charge source is handled numerically by a new method that rids the equations of singularities through a change of dependent variable.  The resulting algorithm is fast compared to earlier methods, works for short bunches with complicated structure,  and yields all six field components at any space-time point.  As an example we compute the tangential magnetic field at the walls. From that one can make a perturbative treatment of the Poynting flux to estimate the energy deposited in resistive walls. The calculation
was motivated by a design issue for LCLS-II, the question of how much wall heating from CSR occurs in the last bend of a bunch compressor and the following straight section. Working
with a realistic longitudinal bunch form of r.m.s. length  $10.4~\mu$m and a charge of 100 pC we conclude that the radiated power is quite small (28 W at a 1 MHz repetition rate), and all radiated energy is absorbed in the walls within 7 m along the straight section.
\end{abstract}  \maketitle

\section{Introduction}
Effects of coherent synchrotron radiation (CSR) need to be studied in almost all advanced accelerator projects. In single-pass systems, for instance Free Electron Lasers, the effects are usually deleterious, for instance in causing transverse emittance degradation in bunch compressors. In electron storage rings CSR may cause unwanted bunch instabilties, but may also be useful in providing intense radiation in the THz domain. In spite of ambitious efforts to provide models and computational tools to describe CSR, there is still a lot of room for improvements in models and algorithms. There are several distinct aspects to the problem: (i) modeling the charge/current distribution in the bunch; (ii) modeling the vacuum chamber; (iii) computing fields of very short bunches which may have micro-structures;
(iv) providing all the field components necessary to describe the experimental situation.

We believe that the paraxial method in the frequency domain offers opportunities for improvements in most of these aspects. It was introduced for the CSR problem in 2003-2004 by Stupakov and Kotelnikov \cite{stupakov-kotelnikov-1} and Agoh and Yokoya \cite{agoh},~\cite{agoh-yokoya}, and pursued since then by Stupakov and Kotelnikov \cite{stupakov-kotelnikov-2}, Zhou \cite{zhou}, Zhou {\it et al.} \cite{zhou_etal}, and Bizzozero {\it et al.} \cite{FEL14},~\cite{FEL13},~\cite{david-thesis}.  An outstanding feature of the method is that it works
better as the frequency, equivalent to wave number $k$, is increased. This has roots in a relation to the ray picture of optics, and provides a chance to
study the fields of very short bunches with possible interior micro-bunching.  Of course, one has to demonstrate the practicality of discretizing in $k$ over a wide range, and taking the inverse Fourier transform to construct fields in space-time.  In the present paper these steps are shown to be entirely practical for bunch parameters close to the current state of the art, namely for a 10 $\mu$m bunch with a realistic longitudinal profile for a section of LCLS-II. We could go to even smaller bunch lengths and account for micro-bunching as well. Thus we have a response to item (iii) above.

For item (ii), the vacuum chamber, most of the cited papers assume a rectangular chamber with perfectly conducting walls. An exception is Ref.\cite{stupakov-kotelnikov-1}, which treats a round chamber. On the other hand, codes that aspire to the realistic simulation of beams usually model the vacuum chamber by infinite parallel plates, if a vacuum chamber is included at all. This might be because the paraxial codes with the rectangular model are thought to be too slow to apply in tracking of macroparticles. We show that the
solution of the paraxial equations with the rectangular model can be speeded up greatly by a more efficient discretization in the transverse coordinates $(x,y)$.
Instead of a finite difference or finite element representation of derivatives in $(x,y)$-space \cite{agoh},~\cite{agoh-yokoya},~\cite{zhou},~\cite{zhou_etal},~\cite{FEL13}, or a two-dimensional mode expansion
\cite{stupakov-kotelnikov-2}, we make a Fourier mode development in the vertical coordinate $y$, treating only $x$ by finite differences.  The motivation for this goes back
to early work \cite{warnock-morton},~\cite{ng}, in which it was noticed that the Fourier series in $y$ converges quickly and also affords
a way to enforce boundary conditions on the horizontal walls. In fact, bringing in the $y$-expansion establishes a fruitful connection to the soluble model of a complete
circular torus and its whispering gallery modes. This connection shows the proper way to enforce low frequency cutoffs at mode-dependent ``shielding thresholds".

Because we avoid
a mode expansion in $x$ it is possible to treat a chamber of varying width in the $x$-direction, provided that variations are not too extreme. That treatment, motivated by our study of a flared chamber at the Canadian Light Source (CLS) \cite{prl}, will be covered in a later report.

Another requirement for the CLS study was to find transverse electric field components at the location of a detector far off the beam axis. Off-axis fields are also needed in the present work to treat resistive wall heating. Thus we have examples of the requirement (iv) above, to find various field components at any point in the chamber.  This requirement is met very easily in the present framework, because the Fourier transforms of all field components are expressed in terms of those for the $y$-components of $\bf E$ and $\bf H$.
After these two field components are computed by solving independent paraxial wave equations it costs almost nothing to find the other four. In particular, the transverse Lorentz force on the beam is readily available.

Item (i), the description of the source, is a crucial step in any scheme and deserves the closest attention. We look forward to a self-consistent scheme in which the bunch is modified by the field it produces, ideally through the Vlasov equation  integrated by the method of local characteristics.
In a less costly approach a macroparticle method replaces the Vlasov description. In any event we want at least a two dimensional representation of the bunch in $(s,x)$ space, where $s$ is the longitudinal coordinate, in order to study bunch compression. As in the cited papers \cite{stupakov-kotelnikov-1}-~\cite{FEL14} we here consider only a one-dimensional description, but in such a way as to clear the path toward higher dimensions. Our charge density has the factored form $q\lambda(s-\beta ct)\delta(x)H(y)$,
where the vertical distribution $H(y)$ is fixed and has a finite width. Other authors put $H(y)=\delta(y)$, which implies an infinite field at $x=y=0$, and so are led to special arguments to extract the relevant finite part. It turns out that the finite width of $H(y)$ is essential in constructing field components other than the longitudinal electric.
 This is a new story, which must be understood prior to attempting a full theory without factorization of the charge distribution.

 In another important departure from earlier work we found an efficient way to deal numerically with  $\delta(x)$ in the source for the wave equation. By a simple change of dependent variable, the wave equation acquires a new effective source in which $\theta(x)$, the unit step function, appears instead of $\delta(x)$. Successive transformations can make the effective source arbitrarily smooth. We find that two transformations, for an effective source proportional to $x\theta(x)$, produces good results. There is a wider scope for this idea, since it can be generalized to smooth and broaden an $x$-distribution which is narrow but not a delta function. This would seem to be the proper approach for a self-consistent scheme describing a low emittance beam, better than the obvious idea of devoting more mesh points to a region near $x=0$.

 It seems clear that our methods could provide a relatively fast  self-consistent scheme with macroparticles. The cost of a competing method, which computes
 fields in space-time through retarded potentials and uses the parallel plate model, is strongly dominated by field evaluations rather than charge/current density construction and particle pushing \cite{bassi}. At each point of observation of the wake field a separate integral over histories is performed, which is very costly. Our method provides a markedly faster field evaluation, while being the same for the other operations. It requires two Fourier transforms to go back and forth between space-time and the frequency domain, but those can be done as FFT's and should not be costly.

 The use of retarded potentials has been revived recently by Stupakov and Zhou \cite{SZ-new}, who find the longitudinal impedance for a line charge source
 in various idealized cases, assuming a parallel plate vacuum chamber. The impedance is in terms of double integrals, within a sum over image charges, which must be evaluated numerically. The local wake field is not obtained, only its integral over time. Our method generalizes directly to a higher dimensional charge distribution, and is more general even for a line charge in that it provides the local wake and allows the  bunch profile to
 vary with $s$.

Our perturbative calculation of the Poynting vector for resistive wall heating follows an established idea \cite{landl}: replace $\bf H$ by its value ${\bf H}_0$ for perfectly conducting boundaries, and replace $\bf E$ by its value from the resistive wall boundary condition, which is approximated by again using ${\bf H}_0$ instead of $\bf H$. As far as we know that scheme is always stated for single modes. Since we have a mode expansion  in only one coordinate, we have to derive new formulas. On the horizontal walls
we find interference between different vertical modes.

Section and subsection titles provide a synopsis of the paper. We include an appendix on the derivation of the resistive wall boundary condition, hoping to clarify certain issues that are not emphasized in textbook treatments \cite{landl}, \cite{jackson}, \cite{dome}.  In particular, we examine the basic assumption that variation of fields within the wall material is primarily in the direction normal to the wall.

\section{ Wave Equation for the Slowly Varying Amplitude in Accelerator Coordinates}
\subsection{ Fourier Transforms in Time and the $y$ Coordinate}
 We work in standard accelerator coordinates (Frenet-Serret coordinates) defined in terms of a reference trajectory $\Rbo(s)$ lying in a plane, and parametrized  by its arc length $s$.  Any spatial point in the laboratory system is represented as $\Rb=\Rbo(s)+x\nb(s)+y\mathbf{e}_y$ where $\nb(s)$ and $\mathbf{e}_y$ are unit vectors
  normal to the unit tangent $\tb(s)=\Rbo\pr(s)$. To be definite we take the horizontal unit vector to be $\nb=\mathbf{e}_y \times \tb$.
 The vacuum chamber is to have a rectangular cross section with planar surfaces at $y=\pm g$, thus with full height $h=2g$. The vertical walls
 at
 \be
 x=x_-\ ,\quad \quad x=x_+   \label{xwalls}
 \ee
 are either planar or cylindrical with constant radius of curvature, depending on $s$.  This accommodates a beam centered at $x=y=0$,
 following a sequence of straights and bends.

Let $F(s,x,y,t)$ be any one of the six field components or a component of the charge/current density. Suppressing for the moment the variables $(x,y)$ we write $F$ as the Fourier integral
\be
F(s,t)=\int_{-\infty}^\infty e^{ik(s-\beta ct)}\hat F(k,s)dk\ ,\label{ft}
\ee
where $\beta c$ is the velocity of the centroid (mean charge position) of the longitudinal charge distribution.
By the Fourier inversion theorem

\be
e^{iks} \hat F(k,s)=\frac{1}{2\pi}\int_{-\infty}^\infty e^{ik\beta ct}F(s,t)\beta cdt\ ,  \label{tft}
\ee
which may also be written as an inversion with respect to $z=s-\beta ct$,
\be
\hat F(k,s)=\frac{1}{2\pi}\int_{-\infty}^\infty e^{-ikz}F(s,(s-z)/\beta c)dz\
\ee
This integral certainly converges in the situation we consider, with a single bunch making one pass through the system. At fixed $s$ the field or source is negligible except for times close to the time of passage of the bunch, say when $|s-\beta ct|$ is less than some multiple of the bunch length.

In defining $\hat F$ as the Fourier transform in time divided by $e^{iks}$ we hope to take out the fastest variation in $s$, leaving $\hat F$ as a slowly varying amplitude.
By examining and solving numerically the differential equations for the $\hat F$ we shall show that this ambition can be realized for the parameters of our example.
Then (\ref{ft}) will be a superposition of waves traveling in the positive $s$-direction, with slowly modulated amplitudes. It is also useful  to view (\ref{ft}) as
a description of the field in the beam frame at any fixed $s$, whether or not the amplitude is slowly varying.

 The electromagnetic boundary conditions for perfectly conducting walls are that the tangential component of $\Eb$ and the normal component of $\Hb$ should vanish.
 We shall meet these conditions on the top and bottom walls (that $E_s, E_x, H_y$ should vanish) by making a Fourier development in $y$.  After restoration of $(x,y)$ the development (\ref{ft}) then takes the form
 \be
F(s,x,y,t)=\int_{-\infty}^{~\infty} dk e^{ik(s-\beta ct)}\sum_{p=0}^{\infty}\phi_p^{(i)}(y) \hat F_p(k,s,x)\ .   \label{fty}
\ee
The choice of the trigonometric function $\phi_p^{(i)}(y)$ to meet the boundary conditions at $y=\pm g$ depends on which field or source component is expanded.
We have
\bea
&&\phi_p^{(1)}(y)=\sin(\alf(y+g))\ ,\quad F=E_s,\ E_x, \ H_y, \ J_s, \ J_x,\ \rho\ , \nonumber\\
&& \phi_p^{(2)}(y)=\cos(\alf(y+g))\ ,\quad F=H_s,\ H_x,\ E_y,\ J_y\ .  \nonumber\\
&& \alf=\pi p/h\ , \label{phichoice}
\eea
where $\Jb$ and $\rho$ are the current and charge densities of the beam.
With these choices the Maxwell equations and boundary conditions are satisfied term-by-term in the sums over $p$. This
follows from  orthogonality,
\be
\frac{1}{g}\int_{-g}^{g}\phi_p^{(i)}(y)\phi_q^{(j)}(y)dy = \delta_{ij}\delta_{pq}\  .   \label{orthog}
\ee
\subsection{Transformed Maxwell Equations in Frenet-Serret Coordinates}
Next write the Maxwell equations using the standard expressions for divergence and curl in curvilinear coordinates. The metric tensor is diagonal with diagonal components
\be
(g_s,\ g_x,\ g_y)=(\eta(x,s),\ 1,\ 1)\ ,\quad  \eta(x,s)=1+x\kappa(s)=1+x/R(s)\ ,     \label{metric}
\ee
where $\kappa(s)$ and $R(s)$ are the curvature and radius of curvature of the reference orbit at $s$.
We substitute fields and sources in the form (\ref{fty}) and take the inverse Fourier transforms with respect to $z=s-\beta ct$ and $y$ to obtain the following system (in SI units).
\bea
&&{\rm div}{\bf\ D}=\rho:\nonumber\\
&&ik\es+\ptl_s\es+\ptl_x(\eta\ex)-\alf\eta\ey=\eta Z_o\rh c\ ,\label{dive}\\
&&{\rm div}{\bf\ B}=0:\nonumber\\
&&ik\hs+\ptl_s\hs+\ptl_x(\eta\hx)+\alf\eta\hy=0\ ,\label{divb}\\
&&{\rm curl}{\bf\ E}+\ptl{\bf B}/\ptl t=0:\nonumber\\
&&\ptl_x\ey-\alf\ex-ik\beta Z_o\hs=0\ , \label{curles}\\
&&\eta\alf\es-ik\ey-\ptl_s\ey-ik\beta \eta Z_o\hx=0\ ,\label{curlex}\\
&&ik\ex+\ptl_s\ex-\ptl_x(\eta\es)-ik\beta\eta Z_o\hy=0\ , \label{curley}\\
&&{\rm curl}{\bf\ H}-\ptl{\bf D}/\ptl t={\bf J}:\nonumber\\
&&\ptl_x\hy+\alf\hx+ik\beta\es/Z_o=\js\ , \label{curlhs}\\
&&-\alf\eta\hs-ik\hy-\ptl_s\hy+ik\beta\eta\ex/Z_o=\eta\jx\ , \label{curlhx}\\
&&ik\hx+\ptl_s\hx-\ptl_x(\eta\hs)+ik\beta\eta\ey/Z_o=\eta\jy\ , \label{curlhy}
\eea
where $Z_o=\mu_o c=1/(\epsilon_o c)$ is the impedance of free space. \pagebreak

These equations may be solved algebraically for all field components in terms of $\ey$ and $\hy$ and their derivatives,
yielding the results
\bea
&&\es=-\frac{1}{\gamma_p^2}\bigg[\frac{\alf}{\eta}(ik\ey+\ptl_s\ey)+i\beta kZ_o(\js-\ptl_x\hy)\bigg]\ ,\label{es}\\
&&\ex=-\frac{1}{\gamma_p^2}\bigg[\alf\ptl_x\ey+i\beta kZ_o\big(\jx+\frac{1}{\eta}(ik\hy+\ptl_s\hy)\big)\bigg]\ ,\label{ex}\\
&&Z_o\hs=-\frac{1}{\gamma_p^2}\bigg[-\alf Z_o\big(\jx+\frac{1}{\eta}(ik\hy+\ptl_s\hy)\big)+i\beta k\ptl_x\ey\bigg]\ ,\label{hs}\\
&&Z_o\hx=-\frac{1}{\gamma_p^2}\bigg[Z_o\alf(\js-\ptl_x\hy)-\frac{i\beta k}{\eta}(ik\ey+\ptl_s\ey)\bigg]\ ,\label{hx}\\
&&\hspace{3 cm}\gamma_p^2=(\beta k)^2-\alf^2\ .\label{g2}
\eea
Here it is assumed that $\gamma_p^2\ne 0$, a condition that is met in our calculations by a $p$-dependent  low frequency cutoff or ``shielding threshold". For a complete treatment allowing arbitrarily low frequencies one can give $k$ a small positive imaginary part.

Equations similar to (\ref{es})-(\ref{hx}) are familiar in a scheme with separate Fourier transforms in $s$ and $t$ \cite{warnock-morton}. Fortunately, with only the one integral transform (\ref{ft}) we can still solve for all fields in terms of $\ey$ and $\hy$ and their derivatives , but a new feature is that $s$-derivatives appear.

 The fields $\ey$ and $\hy$ are obtained as solutions of two independent wave equations with sources. To derive the wave equations one may combine the transformed Maxwell equations as stated above, or proceed from the wave equations in Cartesian form and transform the differential
operator to Frenet-Serret coordinates.  The equation for $\hat F_p=(\hat E_{yp},\hat H_{hp})$ with source $\hat S_p=\big(\hat S_{Ep},\ \hat S_{Hp}\big)$ is
\be
-\frac{1}{\eta^2}\bigg[\bigg(2ik-\frac{\kappa\pr x}{\eta}\bigg)\frac{\ptl\fp}{\ptl s}+\frac{\ptl^2\fp}{\ptl s^2}\bigg]=
\frac{\ptl^2\fp}{\ptl x^2}+\frac{\kappa}{\eta}\frac{\ptl\fp}{\ptl x}+\bigg[\gamma_p^2-\frac{k^2}{\eta^2}
-\frac{ik\kappa\pr x}{\eta^3}\bigg]\fp-\hat S_p\ ,  \label{waveeq}
\ee
where
\be
\hat S_{Ep}=Z_0(\alpha_p c\rh-ik\jy)\ ,\quad\quad \hat S_{Hp}=
\frac{\kappa}{\eta}\js+\frac{\ptl\js}{\ptl x}+\frac{1}{\eta}\big(ik\jx+\frac{\ptl\jx}{\ptl s}\big)\ .    \label{sources}
\ee

The factor $\kappa\pr(s)$ in (\ref{waveeq}) is nonzero where the reference trajectory (which need not be an actual particle trajectory) changes from straight to curved or vice versa. If the change is abrupt at $s=s_0$ then $\kappa\pr(s)$ contains $\delta(s-s_0)$, and it is doubtful that the wave equation can be given a meaning in a neighborhood of that point. On the other hand, if we give $\kappa(s)$
a smooth transition over a distance comparable to a typical fringe field extent in a bending magnet, then each of the terms with
$\kappa\pr$ is small compared to the term immediately preceding it in (\ref{waveeq}).  Accordingly we drop $\kappa\pr$ terms but then
allow $\kappa$ to be a step function at bend-straight transitions, elsewhere in the equation. The solution $\fp$ is required to be continuous at transitions.

Henceforth we drop the transverse currents $\jx,\ \jy$, but these could be restored in a more ambitious self-consistent scheme.

\subsection{Slowly Varying Amplitude Approximation and the Simplified Wave Equation}

We now proceed to the main approximation, which is to assume that the amplitude
$\fp$ in (\ref{fty}) is slowly varying as a function of $s$. We may state
the criterion for slow variation in terms of a norm, for instance
\be
\|f\|=\int_{x_-}^{x_+} |f(x)| dx\ ,  \label{norm}
\ee
where dependence of $f$ on variables other than the transverse coordinate $x$ is suppressed.
Then the requirement on $\hat F_p(k,s,x)$ is
\be
\bigg\|\frac{\ptl^2\fp}{\ptl s^2}\bigg\|\ll 2|k|\bigg\|\frac{\ptl\fp}{\ptl s}\bigg\|\ ,\quad s_0\le s\le s_1\ , \label{sva}
\ee
over the interval of integration $[s_0,s_1]$. This does not make sense as $k\rightarrow 0$, but there is a lower bound to
the relevant $k$ values, as will be shown presently. We shall actually monitor the condition (\ref{sva}) in our calculations,
which apparently has not been done before in similar CSR studies. For convenience (\ref{sva}) is called the Slowly Varying Amplitude (SVA) Approximation or Paraxial Approximation. In our view the former name is more apt, since it reminds us of the only condition that need be enforced in the present framework.

Now within a bend of constant bending radius $R$ the simplified wave equation (\ref{waveeq}) takes the form
\bea
&&\frac{\ptl\fp}{\ptl s}=i\frac{(x+R)^2}{2kR^2}\bigg[\frac{\ptl^2\fp}{\ptl x^2}+\frac{1}{x+R}\frac{\ptl\fp}{\ptl x}
+\bigg(\gamma_p^2-\big(\frac{kR}{x+R}\big)^2\bigg)\fp-\hat S_p\bigg]\ ,\label{bendeq} \\
&&\hat S_{Ep}=Z_0\alpha_p c\rh\ ,\quad\quad \hat S_{Hp}=
\frac{1}{x+R}\js+\frac{\ptl\js}{\ptl x}\ . \label{ssource}
\eea
The corresponding equation in a straight section is obtained in the limit $R \ra \infty$ as
\bea
&&\frac{\ptl\fp}{\ptl s}=\frac{i}{2k}\bigg[\frac{\ptl^2\fp}{\ptl x^2}-\tilde\alpha_p^2\fp-\hat S_p\bigg]\ ,\label{straighteq}\\
&&\tilde\alpha_p^2=\alpha_p^2+k^2/\gamma^2\ ,\quad \hat S_{Ep}=Z_0\alpha_p c\rh\ ,\quad \hat S_{Hp}=\frac{\ptl\js}{\ptl x}\ ,
\eea
where $\gamma$ is the Lorentz factor, $1-\beta^2=1/\gamma^2$.
These equations are sometimes described as ``parabolic", but that is a misnomer. They are of Schr\"odinger type owing to the factor $i$, with mathematical properties different from those of a proper parabolic equation.

Next we choose  a simple factored form for the charge density of the beam, good enough for the present limited study but capable of being generalized. With the corresponding current density it is
\bea
&&\rho(s,x,y,t)=q\lambda(s-\beta ct)\delta(x)H(y)\ ,\quad \Jb(s,x,y,t)=(\ \beta c\rho,\ 0,\ 0\ )\ , \label{rhoj}\\
&& \int\lambda(z)dz=\int H(y)dy=1\ ,\quad q=\int\rho(s,x,y,t)\eta(x,s)dsdxdy\ ,\label{norms}
\eea
where $q$ is the total charge. The continuity equation is satisfied. By (\ref{fty}), (\ref{phichoice}), and (\ref{orthog}), the
Fourier transform with respect to $z$ and $y$ is
\be
\hat\rho_p(k,x)=q\hat\lambda(k)H_p\delta(x)\ ,\quad \hat\lambda(k)=\frac{1}{2\pi}\int e^{-ikz}\lambda(z)dz\ ,\quad
H_p=\frac{1}{g}\int_{-g}^g \sin(\alpha_p(y+g))H(y)dy\ .
\ee
Thus the sources in (\ref{ssource}) become
\be
\hat S_{Ep}=qZ_0\alpha_p c\hat\lambda(k)H_p\delta(x)\ ,\quad \hat S_{Hp}=q\beta c\hat\lambda(k)H_p\big(\delta(x)/R+\delta\pr(x)\big)\ .\label{singsource}
\ee

For an even distribution $H(y)=H(-y)$ we have
\be
H_p=\sin\big(\frac{\pi p}{2}\big)\frac{1}{g}\int_{-g}^g \cos(\alpha_py)H(y)dy\ , \label{evenh}
\ee
so that $H_p$ if zero for even $p$ and alternates in sign for successive odd $p$.
For $H$ we try two examples at opposite extremes in their large-$p$ behavior, a Gaussian of zero mean and r.m.s. deviation $\sigma_y\ll h$, and a square step with the same mean and deviation, thus
\be
H_p=(-1)^{(p-1)/2}~ \frac{1}{g}\exp\big(-\frac{1}{2}(\alpha_p\sigma_y)^2\big)\ , \quad
H_p=(-1)^{(p-1)/2}\frac{1}{g}\frac{\sin(\surd 3\alpha_p\sigma_y)}{\surd 3\alpha_p\sigma_y} \ .\label{hpgsq}
\ee
For the longitudinal distribution we apply the result of a simulation for LCLS-II, to be described presently. A comparison can be made to the Gaussian and square step  cases with formulas
\be
\hat\lambda(k)=\frac{1}{2\pi}\exp\big(-\frac{1}{2}\big(k\sigma_z)^2\big)\ , \quad
 \hat\lambda(k)=\frac{1}{2\pi}\frac{\sin(\surd 3k\sigma_z)}{\surd 3k\sigma_z}\ . \label{lambgsq}
\ee

\section{Numerical Solution of the Simplified Wave Equation}

An elementary way to approach the solution of (\ref{bendeq}) or (\ref{straighteq}) is to discretize the right hand side
on a grid in $x$-space, representing the $x$-derivatives by finite differences. The discretization involves values of the
solution at the boundaries, $\fp(k,s,x_{\pm}(s))$, which are to be fixed at values required by the boundary conditions at a perfect conductor. The equation is then regarded as a system of ordinary differential equations, with $s$ as the independent variable,
in the complex unknowns $\fp(k,s,x_i)\ , i=2,\cdots,N-1$. Here the $x_i$ are the interior points of the $x$-grid. The system
is treated as an initial value problem, the initial value being the $s$-independent solution in an infinite straight wave guide.
\subsection{A Transformation to Mollify the Effective Source}

There is an impediment to discretization, however, due to the $\delta(x)$ and $\delta\pr(x)$ in the source terms (\ref{singsource}).
By a change of the dependent variable, this source can be replaced by a new effective source which behaves as $\theta(x)$ near $x=0$,
where $\theta(x)$ is the Heaviside step function,
 \be
 \theta(x)=0\ ,\ x<0\ ,\quad \theta(0)=1/2\ ,\quad \theta(x)=1\ ,\ x>0\ .\label{thetadef}
 \ee
 A second transformation gives the continuous function $x\theta(x)$ in the effective source, and  successive transformations can make the source arbitrarily smooth.

For $\fp=\ey$ we write the expression in square brackets on the right hand side of (\ref{bendeq}) or (\ref{straighteq}) so as
to emphasize the $x$-dependence, suppressing variables $(k,p)$.  In terms of the differential operator
\be
L=\frac{\ptl^2}{\ptl x^2}+a(x)\frac{\ptl}{\ptl x}+b(x)     \label{Ldef}
\ee
the expression in  square brackets is
\be
\Phi=Lu-c_1\delta(x)\ ,\quad u(x)=\ey(k,s,x) ,\quad c_1=qZ_0\alpha_pc\hat\lambda(k) H_p\ , \label{phi}
\ee
where in the bend
\be
a(x)=\frac{1}{x+R}\ ,\quad b(x)=\gamma_p^2-\big(\frac{kR}{x+R}\big)^2\ ,\label{abdef}
\ee
and in the straight section, with $R=\infty$,
\be
 a(x)=0\ ,\quad  b(x)=-\alpha_p^2-\frac{k^2}{\gamma^2}\ . \label{absdef}
\ee
Now define a new dependent variable $u_1(x)$ by
\be
u(x)=\xi_1(x)+u_1(x)\ , \quad \xi_1(x)=c_1x\theta(x)\ ,\label{xi1def}
\ee
We see that $\xi_1\dpr(x)=c_1\delta(x)$ cancels the original source $c_1\delta(x)$ and we gain a new effective source $S_1$:
\be
\Phi=Lu_1- S_1\ , \quad S_1(x)= -L\xi_1(x)=\sigma_1(x)\theta(x)\ ,\quad \sigma_1(x)=-c_1(a(x)+xb(x))\ . \label{u1eqn}
\ee
Since $\ptl u/\ptl s=\ptl u_1/\ptl s$, the field $u_1$ satisfies the same partial differential equation as $u$, but with
 the new source which is much more suitable for discretization, being piecewise continuous with a jump of $-c_1 a(0)$ at $x=0$.

 Numerical integration of the differential equations with this setup was found to be only partly successful, at least when done by the elementary finite difference method described below. An instability was
  encountered at large $p$ in some cases, a behavior that could be traced to the jump in the source. To remove the jump  we again change the dependent variable to $u_2$ defined by
 \be
 u_1(x)=\xi_2(x)+u_2(x)\ .\quad\xi_2(x) =\sigma_1(0)\frac{x^2}{2}\theta(x)\ .\label{u2def}
 \ee
 This yields a source behaving as $x\theta(x)$ with
\bea
&&\Phi=Lu_2-S_2\ , \quad S_2(x)=-L\xi_2(x)+S_1(x)=\sigma_2(x)x\theta(x)\ ,\quad \nonumber\\
&&\sigma_2(x)=\frac{1}{x}\big(\sigma_1(x)-\sigma_1(0)\big)-\sigma_1(0)\big(a(x)+\frac{x}{2}b(x)\big)\nonumber\\
&& \hskip 1.1cm =c_1\big[\frac{1}{x}\big(a(0)-a(x)\big)+a(0)\big(a(x)+\frac{x}{2}b(x)\big)-b(x)\big]\
\eea

It is clear that this process can be continued for additional smoothing. At the $n$-th stage the source is
\be
 S_n(x)=\sigma_n(x)\frac{x^{n-1}}{(n-1)!}\theta(x)\ .
 \ee
It is determined from $S_{n-1}$ through the transformation
\be
 u_{n-1}=\xi_n(x)+u_n(x)\ ,\quad \xi_n(x)=\sigma_{n-1}(0)\frac{x^n}{n!}\theta(x)\ ,\quad
\ee
thus
\be
S_n=-L\xi_n+S_{n-1}=-\sigma_{n-1}(0)\bigg[\frac{x^{n-2}}{(n-2)!}+a(x)\frac{x^{n-1}}{(n-1)!}+b(x)\frac{x^n}{n!}\bigg]\theta(x)
+\sigma_{n-1}(x)\frac{x^{n-2}}{(n-2)!}\theta(x)\ ,
\ee
hence
\be
\sigma_n(x)=\frac{n-1}{x}\big[\sigma_{n-1}(x)-\sigma_{n-1}(0)\big] -\sigma_{n-1}(0)\big[a(x)+\frac{1}{n}xb(x)\big]\ .\label{sign}
\ee
The divided difference in the first term in (\ref{sign}) is analytic at $x=0$, as a result of $a(x)$ and $b(x)$ being analytic at that point.

A similar procedure works for the magnetic field, even though the source to start with is more singular.
For $\fp=\hy$ the expression in square brackets on the right hand side of (\ref{bendeq}) or (\ref{straighteq}) has the form
\be
\Phi=Lu-c_2(\delta(x)/R+\delta\pr(x))\ ,\quad u(x)=\hy(k,s,x) ,\quad c_2=q\beta c\hat\lambda(k) H_p\ . \label{psi}
\ee
The first transformation to remove $\delta\pr$ also removes $\delta$ because of the special form of $a(x)$. Thus
\be
u(x)=\xi_1(x)+u_1(x)\ , \quad \xi_1(x)=c_2\theta(x)\ ,\label{vdefh}
\ee
yields
\bea
&&\Phi=Lu_1+c_2\big[b(x)\theta(x)+\big(a(x)-1/R\big)\delta(x)\big]\nonumber\\
&&=Lu_1- S_1\ , \quad  S_1(x)=-c_2b(x)\theta(x)\ . \label{veqnh}
\eea
The $\delta$ drops out because $a(0)-1/R=0$ in both the bend and the straight.
As in the discussion above, the second transformation will be
\be
u_1(x)=\xi_2(x)+u_2(x)\ ,\quad \xi_2(x)=-c_2b(0)\frac{x^2}{2}\theta(x)\ ,  \label{wdefh}
\ee
and so on.

Now let us summarize the net effect of two smoothing transformations, invoking the explicit forms of the coefficients $a$ and $b$.
The smoothed field  $u=u_1+u_2=(u_E,u_H)$ is added to $\xi=\xi_1+\xi_2=(\xi_E,\xi_H)$ to make the full field $(\ey,\hy)$, and the effective source is denoted by $\tilde S_p=(\tilde S_{Ep},\tilde S_{Hp})$. In the bend,
\bea
&&\ey=\xi_E(x)+u_E(x)\ ,\quad \xi_E(x)= c_1\bigg[1-\frac{x}{2R}\bigg]x\theta(x)\ ,\nonumber\\
&&\tilde S_{Ep}=c_1\bigg[\frac{2}{R(x+R)}-\big(1-\frac{x}{2R}\big)b(x)\bigg]x\theta(x)\ .\label{smoothe}\\
&&\hy=\xi_H(x)+u_H(x)\ ,\quad \xi_H(x)=c_2\bigg[1-b(0)\frac{x^2}{2}\bigg]\theta(x)\nonumber\\
&&\tilde S_{Hp}=c_2\bigg[ -k^2\frac{x+2R}{(x+R)^2}+b(0)\bigg(\frac{1}{x+R}+\frac{1}{2}xb(x)\bigg)\bigg]x\theta(x)\ . \label{smoothh}\\
&&b(x)=\gamma_p^2-\bigg(\frac{kR}{x+R}\bigg)^2\ ,\quad b(0)=-\alpha_p^2-\frac{k^2}{\gamma^2}\ .
\eea
The corresponding formulas for the straight section are  obtained in the limit $R\ra\infty$, noting that $b(x)$ tends to $b(0)$ in the limit:
\bea
&&\xi_E(x)=c_1x\theta(x)\ ,\quad \tilde S_{Ep}=-c_1b(0)x\theta(x)\ ,\label{smoothes} \\
&&\xi_H(x)=c_2\bigg[1-b(0)\frac{x^2}{2}\bigg]\theta(x)\ , \quad \tilde S_{Hp}=\frac{c_2}{2}(b(0)x)^2\theta(x)\ . \label{smoothhs}
\eea

Note that $\ey$ and $\hy$ must be continuous at the transitions between bend and straight, while the corresponding $u_E,\ u_H$ are not continuous. This must be kept in mind in designing the algorithm for $s$-integration.

In (\ref{ex}) and (\ref{hx}) we have the factor $\js-\ptl_x\hy$, where $\js=c_2\delta(x)$ .  Fortunately, the $c_2\delta(x)$ is cancelled
by the term $c_2\ptl_x\theta(x)$ in $\ptl_x\hy$ as given by (\ref{smoothh}). Such a cancellation was noticed long ago in analytical models \cite{warnock-morton}, but a good way to handle it in a numerical context was lacking before the present innovation.
\subsection{Smoothing Transformation for a Beam of Non-Zero Horizontal Extent}
A more realistic charge/current distribution will have a non-zero extent in $x$ but could still be very narrow, for instance $20~\mu$m in our example from LCLS-II.
To handle that case numerically one can generalize the method described above. For an arbitrary source $S(x)$ we wish to transform
$\Phi=Lu-S$. The first transformation will be

\be
u(x)=\xi_1(x)+u_1(x)\ ,\quad \xi_1(x)=\int_{x_-}^x dx\pr\int_{x_-}^{x\pr} S(x\dpr)dx\dpr\ , \label{firstbr}
\ee
hence
\be
\Phi=Lu_1-S_1\  ,\quad S_1(x)=-a(x)\int_{x_-}^x S(x\pr)dx\pr-b(x)\int_{x_-}^x dx\pr\int_{x_-}^{x\pr} S(x\dpr)dx\dpr\ .
\ee
Now only the integral of $S$ appears in the effective source, a smoother and more extended function than $S$ itself. Of course further transformations
could provide additional smoothing, as we have seen.
\subsection{Boundary Conditions at the Vertical Walls}
With perfect conductivity the boundary conditions at the vertical walls are
\be
\ey(k,s,x_{\pm}(s))=\es(k,s,x_{\pm}(s))=0\ ,\quad \hx(k,s,x_{\pm}(s))=0\ .\label{vertbc}
\ee
From (\ref{es}) and (\ref{hx}) we see that these conditions are met if $\ey$ satisfies a Dirichlet condition and $\hy$ a Neumann condition, namely
\be
\ey(k,s,x_{\pm}(s))=0\ ,\quad \ptl_x\hy(k,s,x_{\pm}(s))=0\ .    \label{vertbcc}
\ee
The corresponding conditions on the smoothed fields $u=(u_E,u_H)$ follow from (\ref{smoothe}) and
(\ref{smoothh}):
\bea
&&u_E(x_-)=0\ ,\quad u_E(x_+)=-\xi_E(x_+) ,\label{wbceb}\\
&& \ptl_x u_H(x_-)=0\ ,\quad \ptl_x u_H(x_+)=-\ptl_x\xi_H(x_+)\ .
 \label{wbchb}
\eea
\subsection{Finite Difference Scheme}
We suppose that  the field values are interpolated  by 4th degree polynomials in $x$, and that derivatives are given by differentiating the interpolation. The 4th degree interpolation \cite{as} of a function $f(x)$ on a grid $\{x_j\}_{j=1}^N$ with uniform cell size $\Delta x$ is
\be
f(x)=\sum_{i=-2}^2 L(\xi,i)f(x_j+i\Delta x)+\epsilon\ ,\quad\quad  x=x_j+\xi\Delta x\ , \label{lagrange}
\ee
with Lagrange polynomials
\bea
&&L(\xi,-2)=\frac{1}{24}(\xi^2-1)\xi(\xi-2)\ ,\nonumber\\
&&L(\xi,-1)=-\frac{1}{6}(\xi-1)\xi(\xi^2-4)\ ,\nonumber\\
&&L(\xi,0)=\frac{1}{4}(\xi^2-1)(\xi^2-4)\ ,\nonumber\\
&&L(\xi,1)=-\frac{1}{6}(\xi+1)\xi(\xi^2-4)\ ,\nonumber\\
&&L(\xi,-2)=\frac{1}{24}(\xi^2-1)\xi(\xi+2)\ .\label{lagrcoeffs}
\eea
The error $\epsilon$ is ${\cal O}((\Delta x)^5)$, and is estimated in terms of the 5-th derivative \cite{as}. For evaluation at interior points of the grid
$x=x_k,\ k=3,\cdots,N-2$ we take $j=k$ and $\xi=0$ for centered interpolation, whereas at border points $x=x_k,\ k=1,2,N-1,N$ we take $j=3,3,N-2,N-2$ with $\xi=-2,-1,1,2$, respectively, for the necessary off-center interpolation.

Differentiating (\ref{lagrange}) with respect to $\xi\Delta x$ gives the formulas for derivatives. Define
\be
L_1(\xi,i)=\frac{1}{\Delta x}\frac{\ptl}{\ptl \xi} L(\xi,i)\ ,\quad L_2(\xi,i)=\frac{1}{(\Delta x)^2}\frac{\ptl^2}{\ptl \xi^2} L(\xi,i)\ .
\ee
Now we can write the discretized form of the wave equation (\ref{bendeq}) for $\ey$  as follows, in terms of the smooth field $u_E$:
\bea
&&\frac{\ptl u_E(x_j)}{\ptl s}=i\frac{(x_j+R)^2}{2kR^2}\bigg[D_2(x_j)+\frac{1}{x_j+R}D_1(x_j)
+\bigg(\gamma_p^2-\big(\frac{kR}{(x_j+R)}\big)^2\bigg)u_E(x_j)-\tilde S_{Ep}(x_j)\bigg]\ ,\nonumber\\
&&\hskip 6cm j=2,\cdots,N-1\ ,\label{discreteb}
\eea
where for $m=1,2$ we have
\bea
&&D_m(x_j)=\sum_{i=-2}^2 L_m(0,i)u_E(x_j+i\Delta x)\ ,\quad j=3,\cdots,N-2\ ,\nonumber\\
&&D_m(x_2)=\sum_{i=-2}^2 L_m(-1,i)u_E(x_3+i\Delta x)\ ,\nonumber\\
&&D_m(x_{N-1})=\sum_{i=-2}^2 L_m(1,i)u_E(x_{N-2}+j\Delta x)\ .   \label{ddef}
\eea
In view of (\ref{wbceb}) the boundary values that appear in these sums are
\be
u_E(x_1)=0\ ,\quad u_E(x_N)=-\xi_E(x_N)\ .   \label{bve}
\ee
where the inner and outer boundaries are at $(x_-,x_+)=(x_1,x_N)$. The equation for $\hy$ in terms of the smooth field $u_H$ has the same form, with the appropriate definitions from (\ref{smoothh}), except that the boundary values
 are expressed in terms of interior values by discretizing the Neumann conditions (\ref{wbchb}):
\bea
&&\ptl_x u_H(x_1)\approx\sum_{i=-2}^2 L_1(-2,i)u_H(x_3+i\Delta x)=0\ ,\nonumber\\
&&\ptl_x u_H(x_N)\approx\sum_{i=-2}^2 L_1(2,i)u_H(x_{N-2}+i\Delta x)=-\ptl_x \xi_H(x_N)\ .\label{magbc}
\eea
Solving for the boundary values we have
\bea
&&u_H(x_1)=-\frac{1}{L_1(-2,-2)}\sum_{i=-1}^2 L_1(-2,i)u_H(x_3+i\Delta x)\ ,\nonumber\\
&&u_H(x_N)=-\frac{1}{L_1(2,2)}\bigg[\sum_{i=-2}^1 L_1(2,i)w_H(x_{N-2}+i\Delta x)+\ptl_x\xi_H(x_N)\bigg]\ . \label{dwhn}
\eea
These values allow the numerical derivatives, as in (\ref{ddef}), to be expressed in terms of interior
values of $u_H$ alone.

In a straight section the discretized equation  (\ref{discreteb}) reduces to.
\be
\frac{\ptl u_E(x_j)}{\ptl s}=\frac{i}{2k}\bigg[D_2(x_j)-\tilde\alpha_p^2u_E(x_j)-\tilde S_{Ep}(x_j)\bigg],\quad
 j=2,\cdots,N-1\ ,\label{discretes}
\ee
with the definitions of (\ref{smoothes}) and the boundary conditions of (\ref{bve}). The equation for $u_H$ is the same
with the definitions (\ref{smoothhs}) and the boundary conditions of  (\ref{dwhn}).

The following calculations will be for a single bend followed by a straight section. The generalization to treat an arbitrary sequence of bends and straights
is obvious, and would not make a great complication in coding because one  has only to specify the curvature $\kappa(s)$ to define the equations at any $s$.
\subsection{Initial Values for the Evolution in  $s$}
The system of linear differential equations (\ref{discreteb}) is to be solved as an initial value problem.  We take the initial value for $s=0$ at the beginning of the bend to be the steady-state field produced by the source in an infinitely long straight chamber. Thus the equation for an initial field $\fp$ is
(\ref{straighteq}) with $\ptl\fp/\ptl s=0$, or
\be
\frac{\ptl^2\fp}{\ptl x^2}-\tilde\alpha_p^2\fp=\hat S_p\ .
\ee
Its general solution is a particular solution plus the general solution of the homogeneous equation,
\be
\fp(x)=A\exp(\tilde\alpha_px)+B\exp(-\tilde\alpha_px)+\int_{x_-}^x \sinh\big(\tilde\alpha_p(x-y)\big)\hat S_p(y)dy\ . \label{geninit}
\ee
in which $A$ and $B$ must be chosen to meet the boundary conditions. With the notation defined in (\ref{phi}) and (\ref{psi}) we have
$\hat S_{Ep}(x)=c_1\delta(x),\ \hat S_{Hp}(x)=c_2(\delta(x)/R+\delta\pr(x))$. Evaluating the integral in (\ref{geninit}) and applying the boundary conditions (\ref{vertbcc}) we find
\bea
&&\ey=\frac{c_1}{\tilde\alpha_p}\bigg[-\frac{\sinh(\tilde\alpha_p x_+)}{\sinh(\tilde\alpha_p(x_+-x_-))}\sinh(\tilde\alpha_p(x-x_-))+
\sinh(\tilde\alpha_px)\theta(x)\bigg]\ ,\label{eyinit}\\
&&\hy=~c_2\bigg[-\frac{\sinh(\tilde\alpha_p x_+)}{\sinh(\tilde\alpha_p(x_+-x_-))}\cosh(\tilde\alpha_p(x-x_-))+
\cosh(\tilde\alpha_px)\theta(x)\bigg]\ . \label{hyinit}
\eea
Note that the definition $\theta(0)=1/2$ in (\ref{thetadef}) makes $\hy(0)=0$ in the case of a centered beam  ($x_+=-x_-$).
For the numerical work it is essential that the definition of $\theta(0)$ be the same in the initial condition as in the
smoothing transformation.

The corresponding initial values of the other field components are derived from (\ref{es})-(\ref{hx}):
\bea
&&\es=\frac{ik}{\alpha_p\gamma^2}\ey\ ,\label{esinit}\\
&&\ex=\frac{Z_0}{\beta}\hy\ ,\label{exinit}\\
&&\hs=0\ , \label{hsinit}\\
&&\hx=-\frac{\beta}{Z_0}\bigg[1+\bigg(\frac{k}{\gamma\gamma_p}\bigg)^2\bigg]\ey\ , \label{hxinit}
\eea
where $\gamma$ is the Lorentz factor. The mechanism for the expected small value of $\es$ at large $\gamma$ (in accord with the familiar disk-like
picture of the field pattern) is the near
cancellation of the terms from $\ey$ and $\js-\ptl_x\hy$ in (\ref{es}). The cancellation becomes  less precise during
field evolution in the bend, but $\es$ is still a small difference of two large terms.

A numerical difficulty arises in the application of (\ref{eyinit}) and (\ref{hyinit}) because of a close cancellation
of large terms at large $x\approx x_+$. The increasing part of the second term in (\ref{eyinit}) or (\ref{hyinit}), namely
$\exp(\tilde\alpha_p x)/2$, cancels against a part of the first term. By some rearrangement we take out the cancelling terms and find the following formulas, suitable
for numerical evaluation:
\bea
&&\ey=-\frac{c_1}{2\tilde\alpha_p}\big[\exp(-\tilde\alpha_px)+\exp(\tilde\alpha_px) (a_1+a_3+a_1a_3)\big]\ ,\nonumber\\
&&\hy=-\frac{c_2}{2}\big[-\exp(-\tilde\alpha_px)+\exp(\tilde\alpha_px) (a_2+a_3+a_2a_3)\big]\ ,\nonumber\\
&& a_1=-\exp(-2\tilde\alpha_px_+)-\exp(-2\tilde\alpha_p(x-x_-))+\exp(-2\tilde\alpha_p(x+x_+-x_-)) \ ,\nonumber\\
&& a_2=-\exp(-2\tilde\alpha_px_+)+\exp(-2\tilde\alpha_p(x-x_-))+\exp(-2\tilde\alpha_p(x+x_+-x_-)) \ ,\nonumber\\
&& a_3=\exp(-2\tilde\alpha_p(x_+-x_-))/(1-\exp(-2\tilde\alpha_p(x_+-x_-)))\ . \label{initfixed}
\eea
Even after this step one must take care to avoid overflow or underflow in evaluation of the exponentials, by appropriate expansions.
\subsection{Evolution in $s$}
Suppressing irrelevant variables we write the system of differential equations for evolution of $u=u_E$ or $u=u_H$ as
\be
\frac{du}{ds}=f(u,s)\ ,  \label{ODE}
\ee
where $u$ and $f$ are vectors with $N-2$ complex components, and $f$ is linear in $u$. For the approximation at $s=s^{(n)}=n\Delta s+s^{(0)}$
we write $\un\approx u(s^{(n)})$, where the integration step $\Delta s$ is allowed to be different in bends from what it is in straight sections. We adopt the leapfrog integration rule, based on the central difference approximation to the derivative:
\be
\frac{u^{(n+1)}-u^{(n-1)}}{2\Delta s}=f(u^{(n)},s^{(n)})\ ,\quad n=1,2,\cdots \ . \label{leapfrog}
\ee
To define $u^{(1)}$ for the first step we use Euler's rule,
\be
\frac{u^{(1)}-u^{(0)}}{\Delta s}=f(u^{(0)},s^{(0)})\ . \label{euler}
\ee
As remarked above, the value of $u$ at the end of a bend is not in general equal to the value of $u$ at the beginning of a following straight, owing to a change in definition of $u$ through source smoothing. Consequently, we use an Euler step to initialize a leapfrog integration in the straight, with the appropriate initial value defined by continuity of the physical (unsmoothed) field at the
bend-straight transition.

Of course there are more powerful methods than the finite difference method for discretizing in $x$ and the leapfrog method for $s$.
We have chosen these simple schemes merely to make our strategies clear and to avoid complications in programming for this exploratory study. We have in fact compared results from a more sophisticated $x$-discretization using the Discontinuous Galerkin Method \cite{hesthaven}, \cite{FEL14}, \cite{david-thesis}, as will be reported below. Future work should look for a method with a good compromise between speed and accuracy.
\section{Poynting Flux at the Walls to Lowest Order}
The Poynting vector $\Eb\times\Hb$ evaluated at a wall describes, through its outwardly directed normal component, the flow of energy into that wall, per unit area and per unit time. At a perfectly conducting wall $\Eb$ is normal to the wall while $\Hb$ is tangential, so the normal component of the Poynting vector vanishes. The resistive wall boundary condition (\ref{finalbc}) implies a tangential component of $\Eb$ at the wall and a non-zero energy flow. We can calculate this flow to lowest order from a knowledge of $\Hb_0$,
the magnetic field computed for perfectly conducting walls. We replace $\Hb$ by $\Hb_0$ in both the second factor of the Poynting vector and in the boundary condition.
In this approximation the Poynting vector $\Sb=\Eb\times\Hb$ at a point $\rb=(s,x,y)$ on the wall  is
\be
\Sb(\rb,t)= (1-i)\bigg(\frac{\beta Z_0}{2\sigma}\bigg)^{1/2}\int dk e^{ik(s-\beta ct)}k^{1/2}\nb\times\hat\Hb_0(k,\rb)
 \times\int dk\pr e^{ik\pr(s-\beta ct)}\hat\Hb_0(k\pr,\rb)\ .\label{ehwall}
\ee
From here on we write $\hat\Hb$ for $\hat\Hb_0$ in accord with the notation of previous sections.

Since we are interested in the total energy loss we may integrate over $t$.
 Note that
\be
\int_{-\infty}^\infty dt\exp(-i\beta c t(k+k\pr))=\frac{2\pi}{\beta c}\delta(k+k\pr)\ ,\quad \hat {\bf H}(-k,\rb)=\hat {\bf H}(k,\rb)^*\ , \label{delrep}
\ee
so that
\be
\int_{-\infty}^\infty \Sb(\rb,t)dt=
(1-i)\bigg(\frac{2Z_0}{\beta\sigma}\bigg)^{1/2}\frac{\pi}{c}\int dk~k^{1/2}\bigg(\nb\times\hat\Hb(k,\rb)\bigg)\times \hat\Hb(k,\rb)^* \ .\label{intpoynt}
\ee
 Here the integrand  has finite support in $t$ because the fields follow the source, and are negligible for $|s-\beta ct|$ greater than some length $L$, the maximum range of wake or predecessor fields. Now notice that
 \be
  \bigg(\nb\times\hat\Hb\bigg)\times \hat\Hb^*=(\nb\cdot\hat\Hb^*)\hat\Hb-(\hat\Hb\cdot\hat\Hb^*)\nb\ = -(\hat\Hb\cdot\hat\Hb^*)\nb\   ,\label{neat}
 \ee
 since $\Hb$ satisfies the boundary condition for a perfect conductor, with zero normal component. Moreover, $(1-i)k^{1/2}$ goes into its complex conjugate as $k\rightarrow -k$, since $k^{1/2}\rightarrow i|k^{1/2}|$ as we have defined it in the complex plane in Appendix A. Then in view of (\ref{delrep}) the integral on $k$ is
 twice the real part of the integral on positive $k$ and
 \be
  \int_{-\infty}^\infty {\bf S}(\rb,t)dt=
-\nb\bigg(\frac{2Z_0}{\beta\sigma}\bigg)^{1/2}\frac{2\pi}{c}\int_0^\infty dk~k^{1/2}\hat\Hb(k,\rb)\cdot\hat\Hb(k,\rb)^* \ .\label{kpos}
 \ee
We see that the time-integrated energy flux is solely along the normal direction and is positive toward the wall at all points (since $\nb$ is directed inward toward the vacuum).

Next we wish to integrate (\ref{kpos}) over one transverse dimension at the walls; namely, over $y$ at $x=x_\pm$ for vertical walls and over $x$ at $y=\pm g$ for horizontal walls. By (\ref{fty}) and (\ref{phichoice}) the Fourier development in $y$ is
\be
\hat H(k,s,x,y)=\sum_{p(odd)=1}^\infty \bigg(\us\phi_p^{(2)}(y)\hs(k,s,x)+ \ux\phi_p^{(2)}(y)\hx(k,s,x)+\uy\phi_p^{(1)}(y)\hy(k,s,x)\bigg)\ . \label{ydevel}
\ee
On the vertical walls this reduces to
\be
 \hat H(k,s,x_\pm,y)=\sum_{p(odd)=1}^\infty \bigg(\us\phi_p^{(2)}(y)\hs(k,s,x_\pm)+\uy\phi_p^{(1)}(y)\hy(k,s,x_\pm)\bigg)\ .  \label{ydevelv}
\ee
while on the horizontal walls it becomes
\be
\hat H(k,s,x,\pm g)=\pm\sum_{p(odd)=1}^\infty \bigg(\us\hs(k,s,x)+ \ux\hx(k,s,x)\bigg)\ . \label{ydevelh}
\ee
At the vertical walls we can use the orthogonality of (\ref{orthog}) to find the $y$-integral as
\bea
&&-\nb\cdot\int_{-g}^g dy\int_{-\infty}^\infty dt\ \Sb(s,x_\pm,y,t)=\nonumber\\
&&\bigg(\frac{2Z_0}{\beta\sigma}\bigg)^{1/2}\frac{2\pi g}{c}\int_0^\infty dk~k^{1/2}\sum_p\bigg(|\hs(k,s,x_\pm)|^2+ |\hy(k,s,x_\pm)|^2\bigg)\ . \label{vheat}
\eea
At the horizontal walls the $x$-integral is
\bea
&&-\nb\cdot\int_{x_-}^{x_+}dx\int_{-\infty}^\infty dt\ \Sb(s,x,\pm g,t)=\nonumber\\
&&\bigg(\frac{2Z_0}{\beta\sigma}\bigg)^{1/2}\frac{2\pi}{c}\int_0^\infty dk~k^{1/2}\int_{x_-}^{x_+}dx
\bigg(|\hat H_s(k,s,x,\pm g)|^2+|\hat H_x(k,s,x,\pm g)|^2\bigg)\ , \label{hheat}
\eea
where
\be
\hat H_s(k,s,x,\pm g)=\sum_p\cos(\alpha_p(\pm g+g)\hat H_{sp}(k,s,x)=\mp\sum_p\hat H_{sp}(k,s,x)\ ,
\ee
with the same formula holding for $\hat H_x$.
To find the total energy deposited in the walls the expressions (\ref{vheat}) and (\ref{hheat}) must be integrated over $s$ using the numerical
solutions  for the tangential $\Hb$ fields.
\section{ Total Energy Radiated and the Wake Field}
Here we derive the formula for the total energy radiated, for comparison to the amount of energy absorbed in resistive walls. By conservation of energy this is just the negative of the work done on the beam by the longitudinal component of the electric field. The work done on an infinitesimal charge element $dQ=\rho(\rb,t)d\rb$ in time $dt$ is
\be
dW=\rho(\rb,t)d\rb E_s(\rb,t)\beta cdt\ ,\quad E_s(\rb,t)=\int dk~e^{ik(s-\beta ct)}\sum_{p(odd)=1}^\infty\sin\alpha_p(y+g)\es(k,\rb)\ . \label{dw}
\ee
It follows that the power radiated from all elements is
\be
P=d{\mathcal E}/dt=-\beta c\int d\rb\rho(\rb,t)E_s(\rb,t)\ ,\label{power}
\ee
and the energy radiated while the bunch center moves from $s=0$ to $s=\bar s$ is
\be
{\mathcal E}(0,\bar s)=-\beta c\int_0^{\bar s/\beta c} dt\int d\rb\rho(\rb,t)E_s(\rb,t)\ , \label{calE}
\ee
For our simple model of the charge density in (\ref{rhoj}) we have
\bea
&&P=-q\beta c\int dsdxdy \lambda(s-\beta ct)\delta(x)H(y)\int dk~e^{ik(s-\beta ct)}\sum_{p(odd)=1}^\infty\sin\alpha_p(y+g)\es(k,s,x)\nonumber\\
&&\hskip 5mm =-q\beta c g\sum_p H_p\int dk \int ds\lambda(s-\beta ct)e^{ik(s-\beta ct)}\es(k,s,0)\ .  \label{Psimp}
\eea
The slowly varying amplitude $\es(k,s,0)$ changes little over the length of the bunch, so that it may be replaced by  $\es(k,\beta ct,0)$ in (\ref{Psimp}).
Thus the $s$-integral gives just the conjugated Fourier transform of $\lambda$ so that
\be
P= -2\pi q\beta c g\sum_p H_p\int dk ~\hat\lambda_k^*~\es(k,\beta ct,0)=-4\pi q\beta c g\sum_p H_p\ \rep\int_0^\infty dk ~\hat\lambda_k^*~\es(k,\beta ct,0)\ ,
\ee
and
\be
 {\mathcal E}(0,\bar s)=-4\pi q\beta c g\sum_p H_p\ \rep\int_0^\infty dk ~\hat\lambda_k^*~\int_0^{\bar s}ds~\es(k,s,0)\ .\label{eofs}
\ee

For comparison to earlier work we are also interested in the longitudinal wake field,
\be
W(z,s,x,y)=2{\rm Re}\int_0^\infty dk~e^{ikz}\sum_p\sin\alpha_p(y+g)\es(k,s,x)\ ,\quad z=s-\beta  ct\ . \label{wakexy}
\ee
We evaluate this  at $x=0$ and and take its mean value with respect to the vertical charge distribution  $H(y)$ to obtain
\be
 W(z,s)=2g{\rm Re}\int_0^\infty dk~e^{ikz}\sum_pH_p\es(k,s,0)\ .\label{wake}
\ee
\section{Numerical Results}
\subsection{Parameters and Bunch Profile for LCLS-II}
\begin{figure}[htb]
   \centering
   \includegraphics*[width=.8\linewidth, height=.4\linewidth]{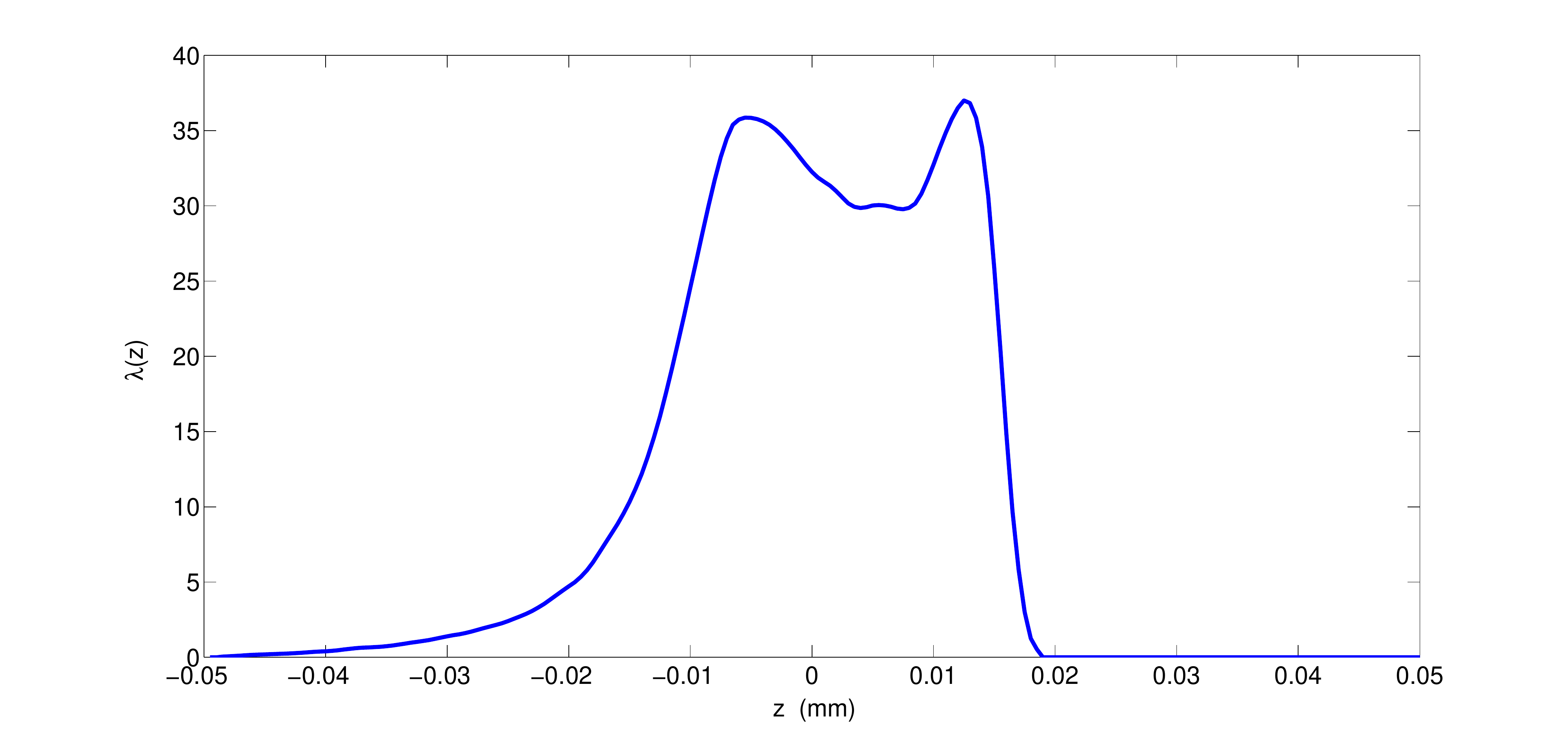}
   \caption{   The simulated form of the LCLS-II bunch at the end of the second bunch compressor. It has r.m.s. length $\sigma_z=10.34\ \mu$m and zero mean. This is a smoothed version of a histogram with 100 bins. }
   \label{fig:bunch}
\end{figure}
\begin{figure}[htb]
   \centering
   \includegraphics*[width=.8\linewidth,height=.4\linewidth]{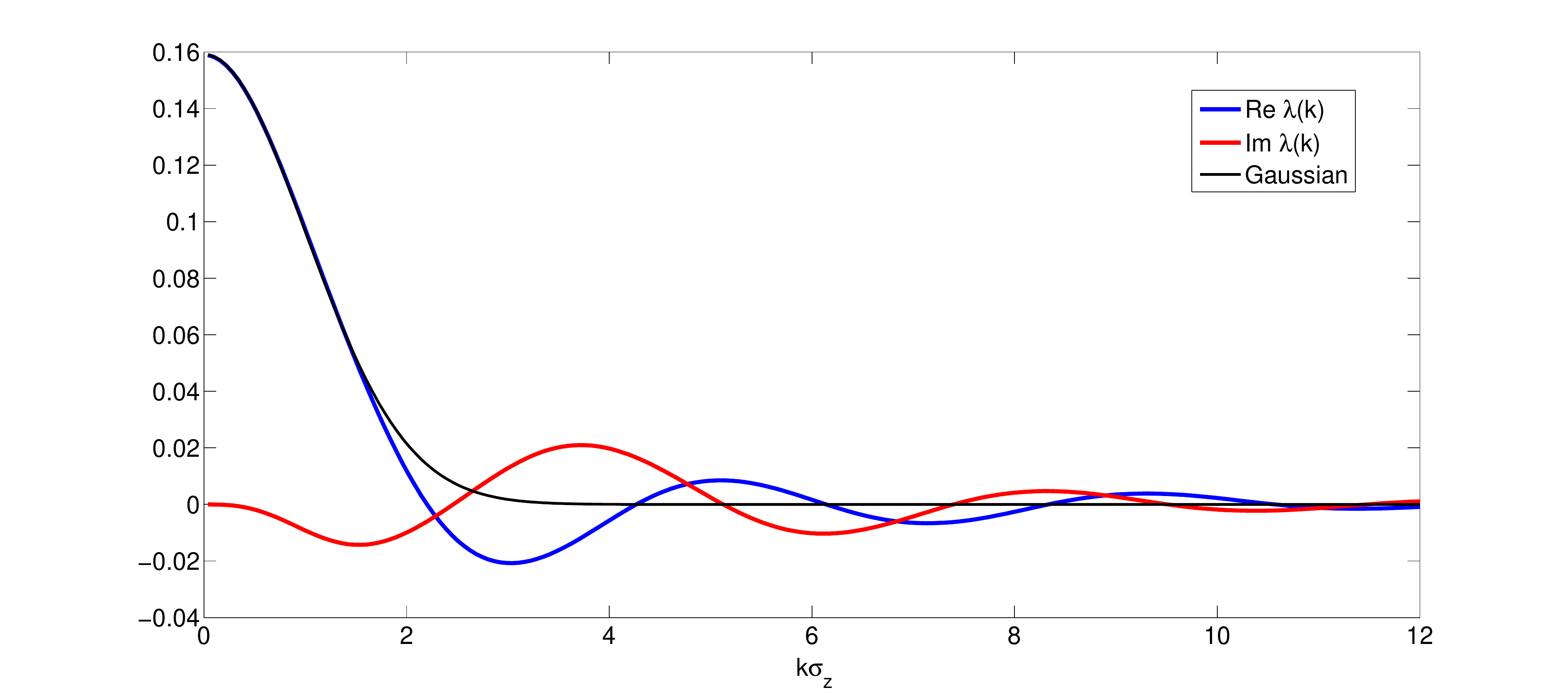}
   \caption{ Fourier transform $\hat\lambda(k)$ of the simulated form of the LCLS-II bunch at the end of the second bunch compressor, compared to that of a Gaussian with the same $\sigma_z$.}
   \label{fig:lambk}
\end{figure}
We present some numerical results, mostly with parameters anticipated for the final bend of the second bunch compressor (BC2) in the forthcoming Linac Coherent Light Source II (LCLS-II) at SLAC National Accelerator Laboratory. The beam is centered in the rectangular chamber of width $w=2x_+$ and height $h$. The chamber dimensions, bending radius, and bend angle are
\be
w=5\ {\rm cm}\ ,\quad h=2 \ {\rm cm}\ ,\quad R=12.9\ {\rm m}\ ,\quad \theta=42.5\ {\rm mrad}\ .\label{para1}
\ee
The chamber material is copper; we take the conductivity to be $\sigma=5.96\cdot10^7\ \Omega^{-1} {\rm m}^{-1}$.
The single-bunch charge, energy, and nominal r.m.s.  beam dimensions are
\be
q=100\ {\rm pC}\ ,\quad E= 1.6\ {\rm GeV}\ ,\quad \sigma_z=0.010\ {\rm mm}\ ,\quad \sigma_x=0.02\ {\rm mm}\ ,\quad \sigma_y=0.16 \ {\rm mm}\ .\label{para2}
\ee
There is also a mode with $q=300~$pC and a longer bunch, which is of less interest for this study.
The repetition rate can be as great as $ 1\ {\rm mHz}$. We calculate the fields and energy loss for a single bunch. In view of the small value of $\sigma_x$
we represent the bunch charge density as in (\ref{rhoj}), with zero width in $x$. For the fields to be finite at the beam it is then necessary that there be a non-zero spread in the vertical density $H(y)$. We compare the Gaussian and square-step distributions with Fourier transforms (\ref{lambgsq}).

 For the longitudinal distribution $\lambda(z)$ we apply the result of a realistic simulation,  and compare the outcome to that for a Gaussian with the same $\sigma_z$.
   A smoothed representation of the  simulated distribution is shown in Fig.\ref{fig:bunch}; it has zero mean. The simulation gave a histogram of 100 bins, which was smoothed by convolving with a quartic kernel having a half-width of 3 bins and zero slope at the ends. The real and imaginary parts of the Fourier transform $\hat\lambda(k)$ of the distribution are shown in Fig.\ref{fig:lambk}, along with the corresponding transform of a Gaussian with the same $\sigma_z$. The smoothing has no discernible effect on the Fourier transform in the range of $k$ plotted.

    We set $\beta=1$, but our code allows $\beta<1$ which is at least of interest for a general understanding of fields and code diagnosis,
if not for immediate applications.

\subsection{Initial Conditions and Convergence of the Vertical Mode Expansion}

\begin{figure}[htb]
   \centering
   \includegraphics*[width=.8\linewidth,height=.4\linewidth]{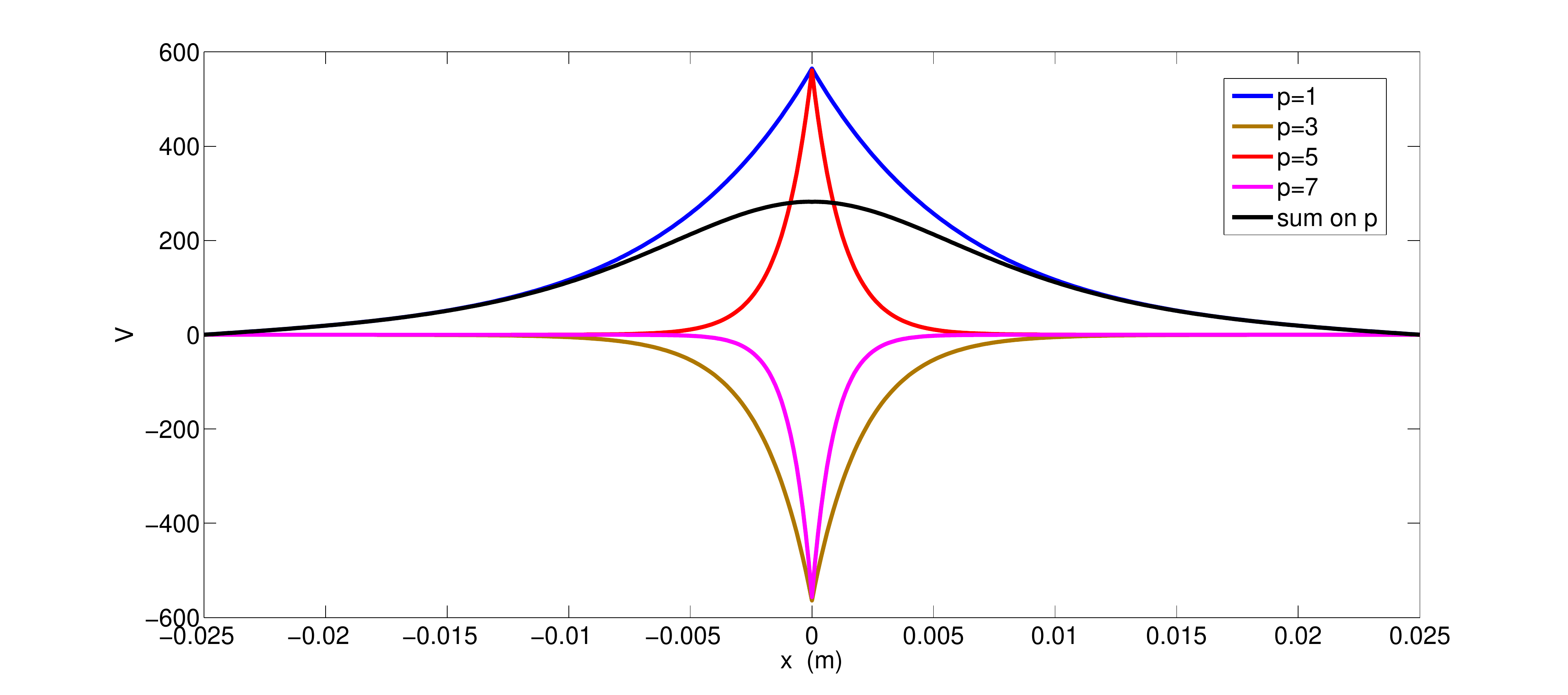}
   \caption{
   The first four modes of the initial field $\ey(x)$ and the sum over all modes at $y=g$. The factor $\hat\lambda(k)$ is omitted. $p=1$ (blue); $p=3$ (brown);
   $p=5$ (red); $p=7$ (magenta); Sum on p (black)}.
   \label{fig:einit}
\end{figure}

We first show the behavior of the vertical field  $\ey(x)\ ,\ \hy(x)$ from which all other field components are derived via Eqs.~(\ref{es})-(\ref{hx}).
The integration on $s$ begins at $s=0$, the beginning of the bend, with the steady state solutions for the straight pipe given in (\ref{eyinit}) and (\ref{hyinit}). In Fig.\ref{fig:einit} we  show the initial $\ey(x)$ for the first four modes ($p=1,3,5,7$) and the sum over all modes,
\be
\hat E_y(k,x,y)=\sum_{p({\rm odd})}\cos\big(\alpha_p(y+g)\big)\ey(k,x)\ , \label{sumeyp}
\ee
evaluated at the upper boundary $y=g$ where the cosine is $-1$. The only $k$-dependence is through the dimensionless factor $\hat\lambda(k)$ which is omitted in the plots.
Convergence of the $p$-sum at $x=0$ happens only by virtue of the  decay of $H_p$, which is not appreciable until large $p$ because of the small value of $\sigma_y$. For the
Gaussian we have convergence (as judged by graphical inspection) by $p=139$, whereas for the square step we go to $p=1999$. The limit is the same for the two cases, evidently because the field far from the beam at $y=g$ is not sensitive to the vertical charge density. The spikes at $x=0$ get narrower with increasing $p$, providing convergence at $x\ne 0$ (but not uniform in $x$) without the help of the factor $H_p$.  Short of the limit there is a narrow spike in the sum, which alternates in direction (up or down) as each new mode is added.
\begin{figure}[htb]
   \centering
   \includegraphics*[width=\linewidth,height=.4\linewidth]{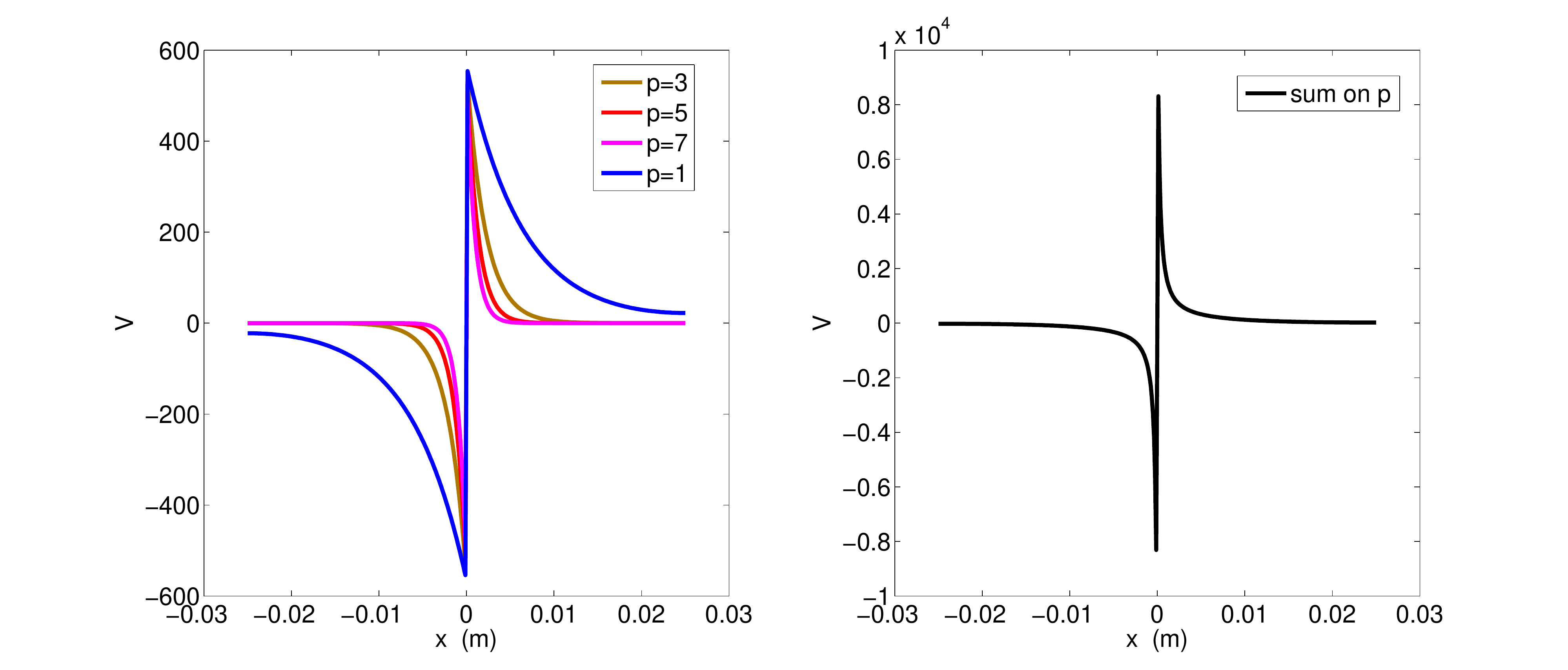}
   \caption{
   Left: the first four modes of the initial field $Z_0\hy(x)$. The factor $\hat\lambda(k)$ is omitted. $p=1$ (blue); $p=3$ (brown);
   $p=5$ (red); $p=7$ (magenta). Right: sum over all $p$ of $Z_0\hy(x)$.}
   \label{fig:hinit}
\end{figure}

Fig.\ref{fig:hinit} (left) shows the first four modes of the initial $\hy$,

\be
\hat H_y(k,x,y)=\sum_{p({\rm odd})}\sin\big(\alpha_p(y+g)\big)\hy(k,x)\ , \label{sumeyp}
\ee
evaluated at $y=0$ where the sine is equal to the alternating factor $(-1)^{(p-1)/2}$. Since $\hy$ has the same alternating factor arising from $H_p$, the summand lacks the
alternating sign that appeared in (\ref{sumeyp}). Consequently the modes just add up coherently, giving the limit shown in Fig.\ref{fig:hinit} (right). Note that we have included the factor $Z_0$ because $Z_0\hy$ and $\ey$ have  the same units (volts) and can be compared in magnitude.
\subsection{Characteristics and Evolution of the Various Field Components}

Next we show the evolution of the initial fields in the bend, for a particular value of the wave number $k$ within the important range of wave numbers.
\begin{figure}[htb]
   \centering
   \includegraphics*[width=.8\linewidth,height=.4\linewidth]{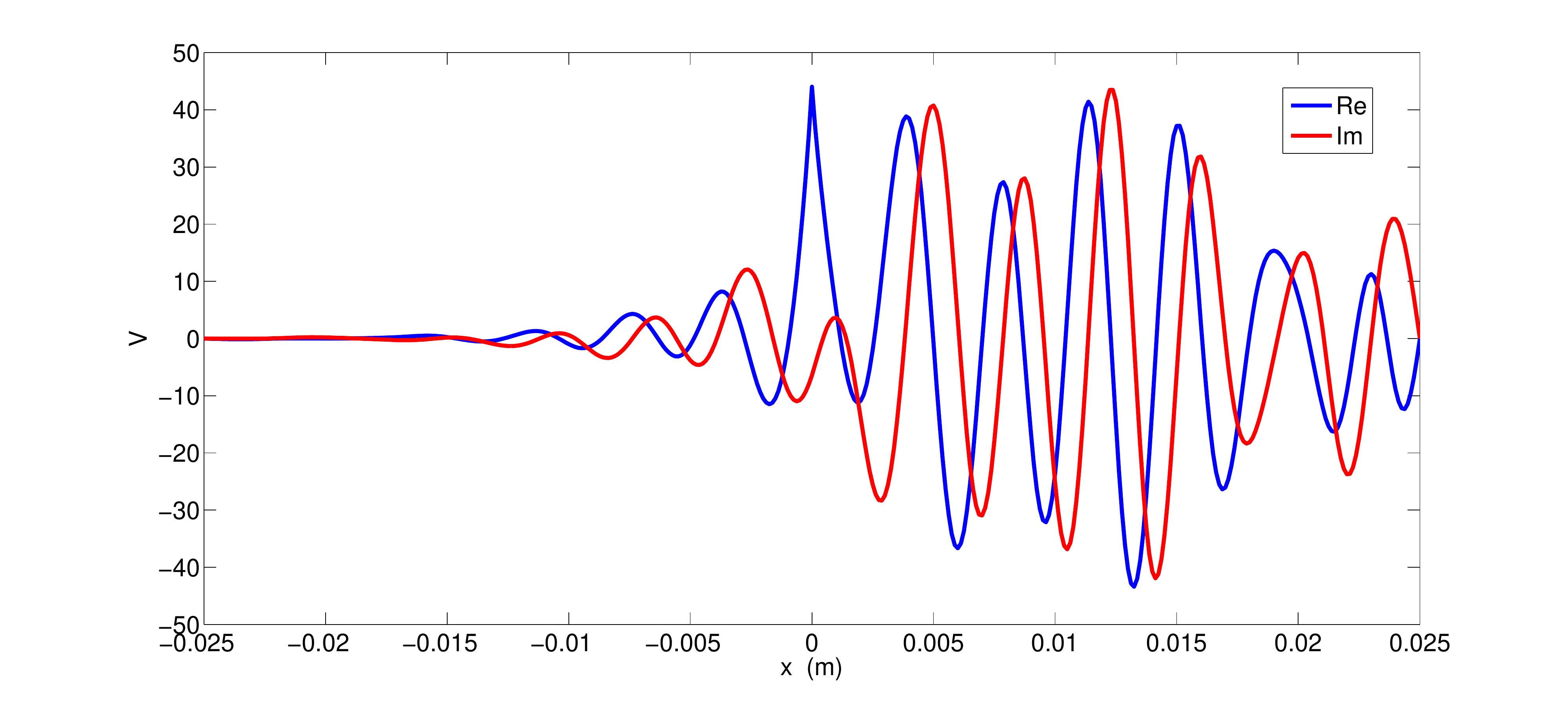}
      \caption{
   The sum to $p=9$ of $\ey(k,s,x)$ at $y=g$, at the end of the bend ($s=0.55~$m) for $kR=5\cdot10^{5}$.}
   \label{fig:eybend5e5}
\end{figure}
\begin{figure}[htb]
   \centering
   \includegraphics*[width=.8\linewidth,height=.4\linewidth]{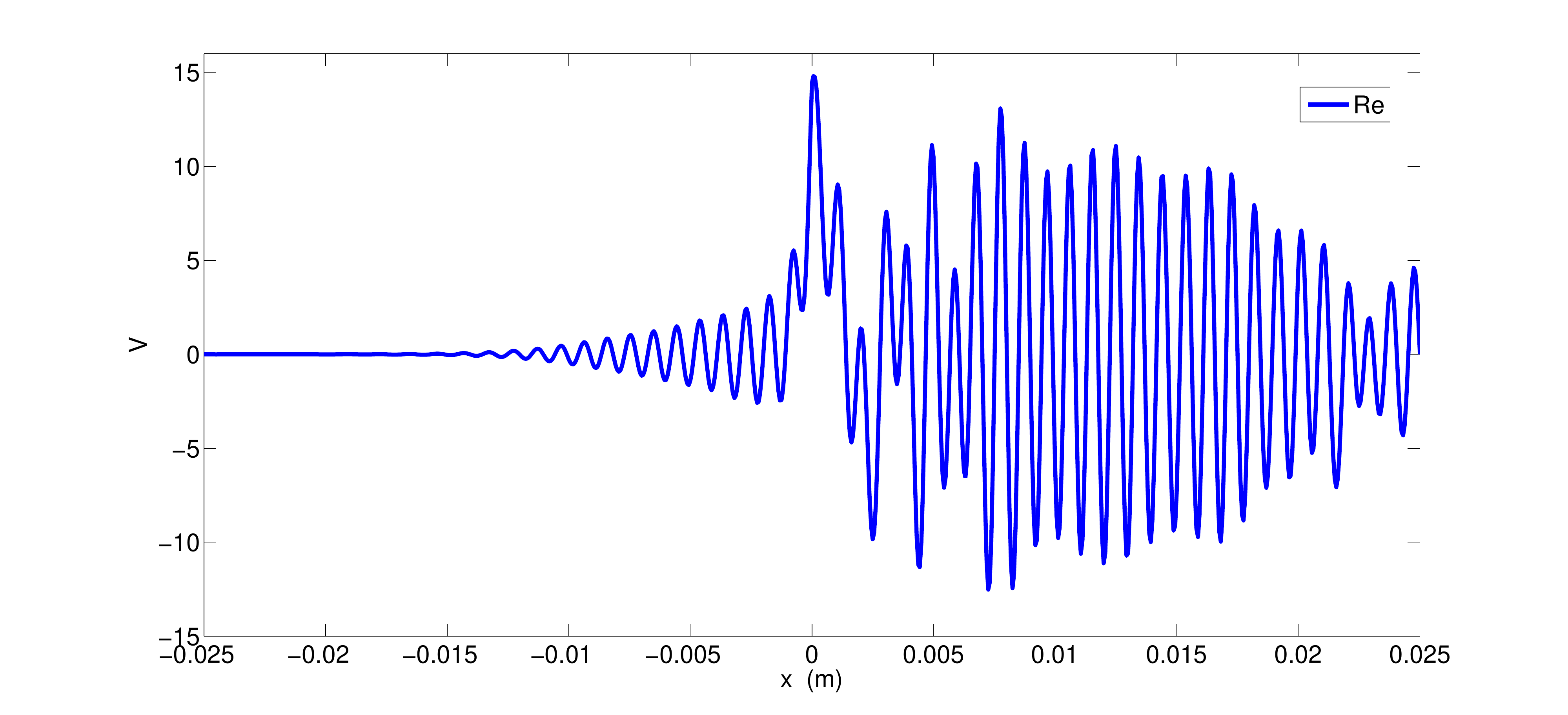}
      \caption{
   The sum to $p=9$ of $\ey(k,s,x)$ (real part) at $y=g$, at the end of the bend ($s=0.55~$m) for $kR=2\cdot10^{6}$.}
   \label{fig:eybend2e6}
\end{figure}
 \begin{figure}[htb]
   \centering
   \includegraphics*[width=.8\linewidth,height=.4\linewidth]{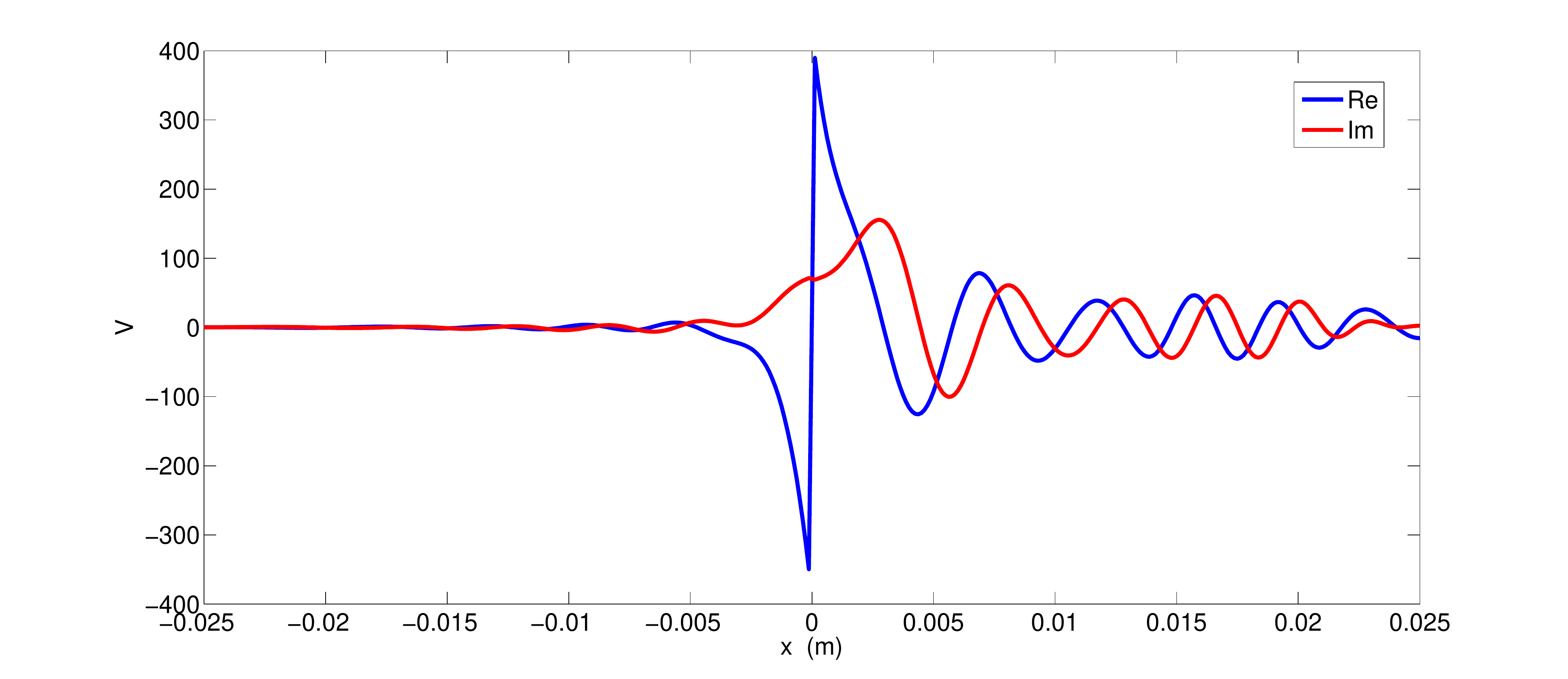}
      \caption{
   The sum to $p=9$ of $Z_0\hy(k,s,x)$  at $y=0$, at the end of the bend ($s=0.55~$m) for $kR=5\cdot10^{5}$.}
   \label{fig:hybend5e5}
\end{figure}
 \begin{figure}[htb]
   \centering
   \includegraphics*[width=.8\linewidth,height=.4\linewidth]{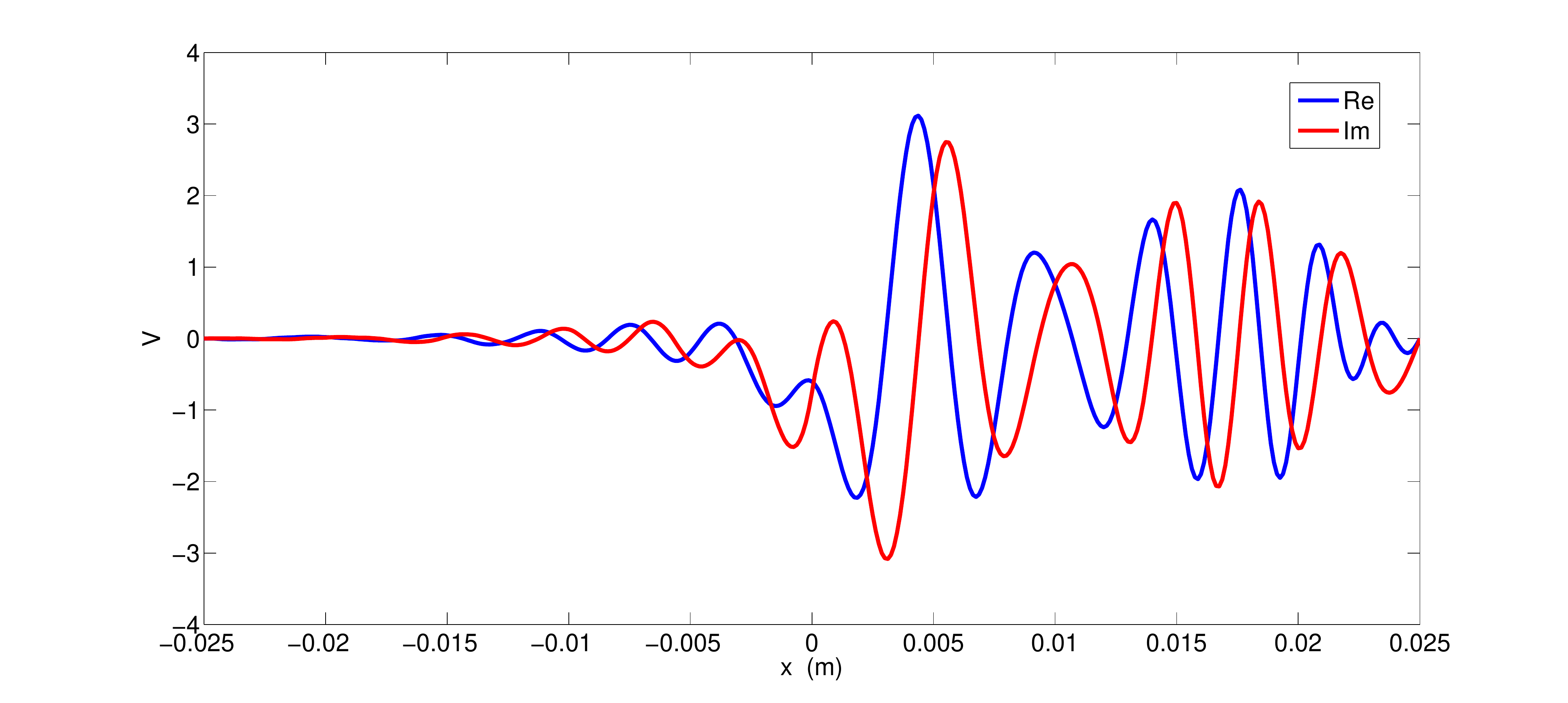}
   \caption{
  The sum to $p=9$ of $\hat E_{sp}(k,s,x)$  at $y=0$, at the end of the bend ($s=0.55~$m) for $kR=5\cdot10^{5}$.}
   \label{fig:esbend5e5}
\end{figure}
 \begin{figure}[htb]
   \centering
   \includegraphics*[width=.8\linewidth,height=.4\linewidth]{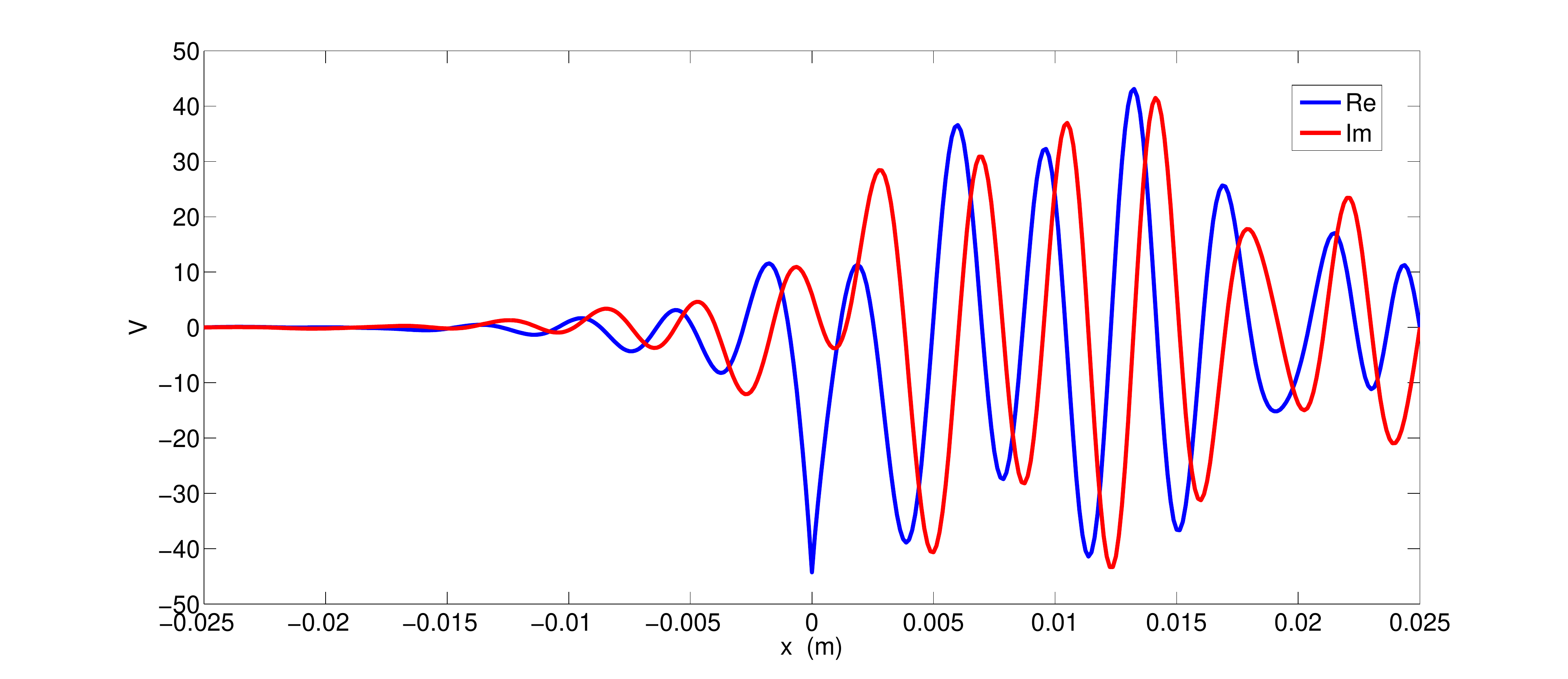}
   \caption{
  The sum to $p=9$ of $Z_0\hat H_{xp}(k,s,x)$  at $y=g$, at the end of the bend ($s=0.55~$m) for $kR=5\cdot10^{5}$.}
   \label{fig:hxbend5e5}
\end{figure}
 \begin{figure}[htb]
   \centering
   \includegraphics*[width=.8\linewidth,height=.4\linewidth]{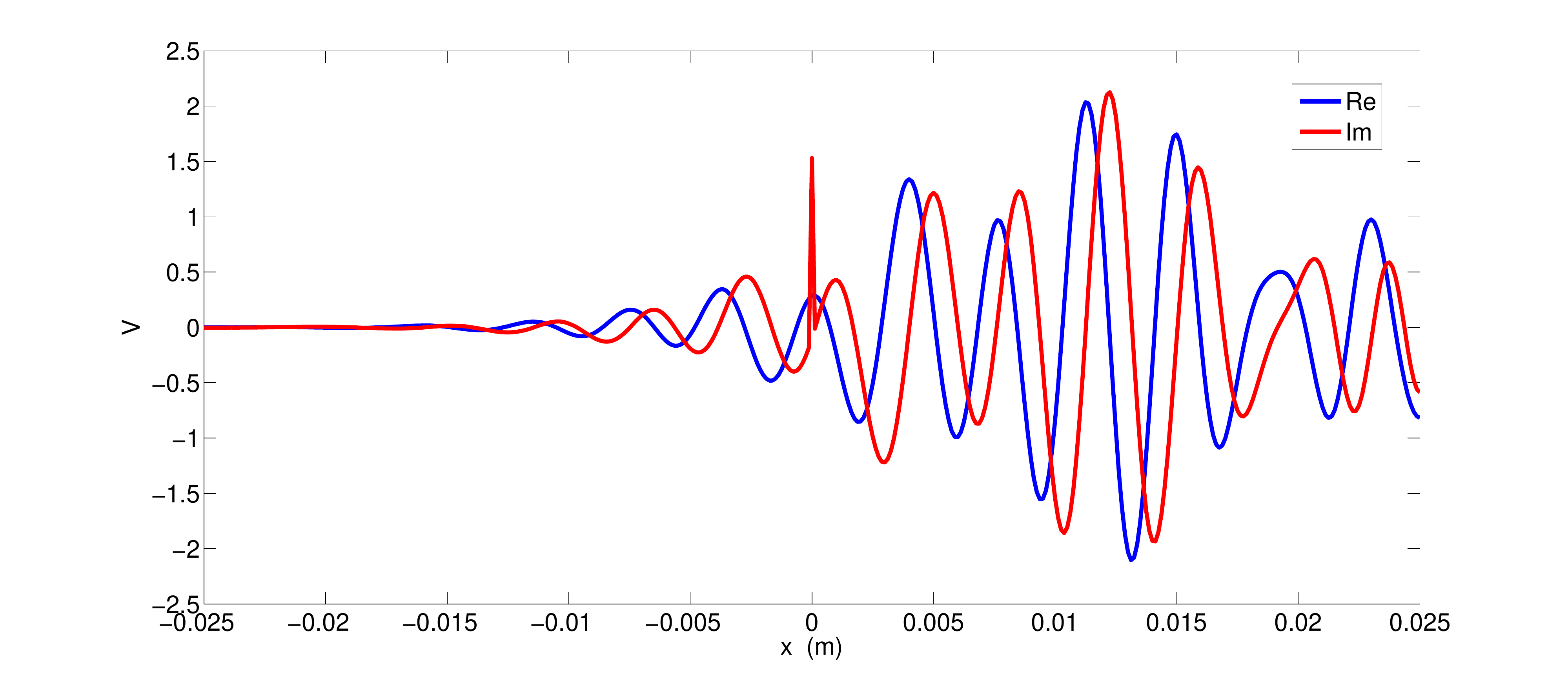}
   \caption{
  The sum to $p=9$ of $Z_0\hat H_{sp}(k,s,x)$  at $y=g$, at the end of the bend ($s=0.55~$m) for $kR=5\cdot10^{5}.$}
   \label{fig:hsbend5e5}
\end{figure}

Fig.\ref{fig:eybend5e5} shows the real and imaginary parts of the sum to $p=9$ of $\ey(k,s,x)$  at $y=g$, for $s$ at the end of the bend and $kR=5\cdot 10^5$. The spike in the real part is inherited from the initial field, and as before it alternates in sign with the number of terms in the sum. As in the initial field, convergence is achieved only at very high $p$. The spike is absent in the imaginary part.

 Fortunately, we can deal with convergence at high $p$ because the high-$p$  modes do not evolve appreciably in the bend and the following straight section. Starting at $p=11$ or so the change of a mode is so small as to be nearly invisible in a graph. Hence our procedure will be to represent the high-$p$ part of $\hat E_y,\ \hat H_y$ and their $x$-derivatives by their given initial values.

 The number of oscillations increases with $k$. Increasing $k$ by a factor of $4$ we get the plot (of the real part) of $\ey$ in Fig.\ref{fig:eybend2e6}.

Next we show graphs for the other field components that enter the calculation, for the same parameters as used in Fig.\ref{fig:eybend5e5}. Fig.\ref{fig:hybend5e5} shows $\hat H_y$ at $y=0$ at the end of the bend. The magnitude of the jump in the real part at $x=0$ will grow as the sum on $p$ is extended, to approach the value seen in Fig.\ref{fig:hinit} (right).

Fig.\ref{fig:esbend5e5}
shows the longitudinal electric field $\hat E_s $ evaluated at $y=0$. It shows no trace of a spike at $x=0$, even though it is constructed from (\ref{es}) with contibutions from $\ey$ and $\ptl_x\hy$, both of which have big spikes at $x=0$. The spikes cancel each other in a way that is certainly remarkable but mathematically obscure.  One can understand the cancellation of high-$p$ contributions, which is the same cancellation that occurs in the initial condition to give a residual of order $1/\gamma^2$ seen in (\ref{esinit}). The cancellation at small $p$ is the puzzling point.  In any event, just a few terms in the $p$-sum are needed to represent $\hat E_s $ and the longitudinal wake field derived from it.

In Fig.\ref{fig:hxbend5e5} we have $Z_0\hat H_x$ evaluated at $y=g$.  In this component the spikes do not cancel, and the behavior of the spikes is like that of $\hat E_y$;
a hint of this can be seen in the initial value (\ref{hxinit}).  In computing the resistive wall loss on the horizontal walls the high-$p$ sum must be included in $\hat H_x$
to get convergence.

Finally in Fig.\ref{fig:hsbend5e5} we show $Z_0\hat H_s$ evaluated at $y=g$.  Now the spike is in the imaginary part, and comes from adding jumps of opposite sign in the terms with $i\hy$ and $i\ptl_x\ey$. Again the spike alternates in sign with the number of terms in the sum on $p$, and convergence is achieved by including the high-$p$ sum. Note that $\hat E_s$ and $\hat H_s$ are relatively small, in accord with their initial values, $\mathcal{O}(1/\gamma^2)$ and $0$, respectively.

The satisfaction of appropriate boundary conditions is apparent in the various graphs.

\begin{figure}[htb]
   \centering
   \includegraphics*[width=.8\linewidth,height=.4\linewidth]{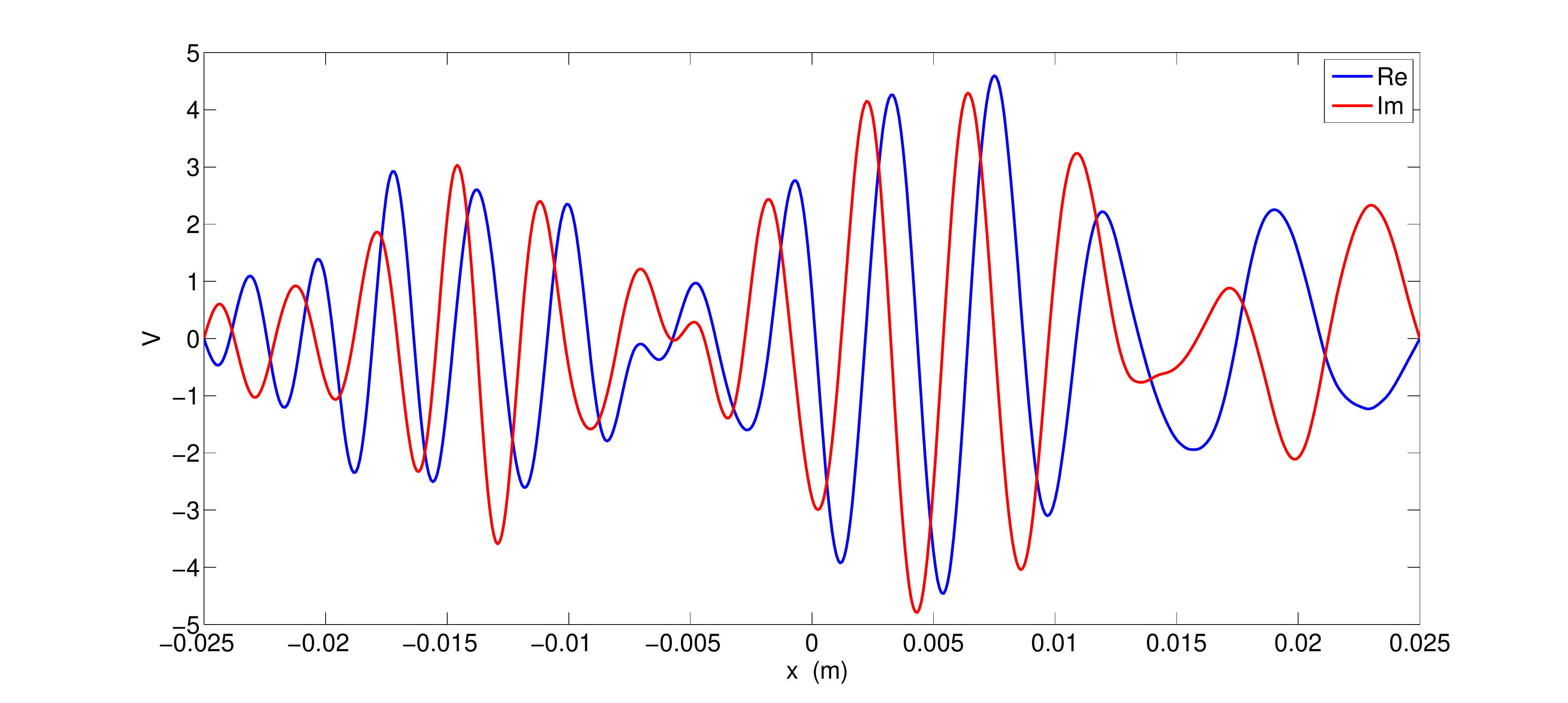}
   \caption{
  The sum to $p=9$ of $\hat E_{sp}(k,s,x)$  at $y=0$, at $1$m into the straight after the bend($s=1.55~$m) for $kR=5\cdot10^{5}.$}
   \label{fig:esbend1mstraight}
\end{figure}

\begin{figure}[htb]
   \centering
      \includegraphics*[width=.8\linewidth,height=.4\linewidth]{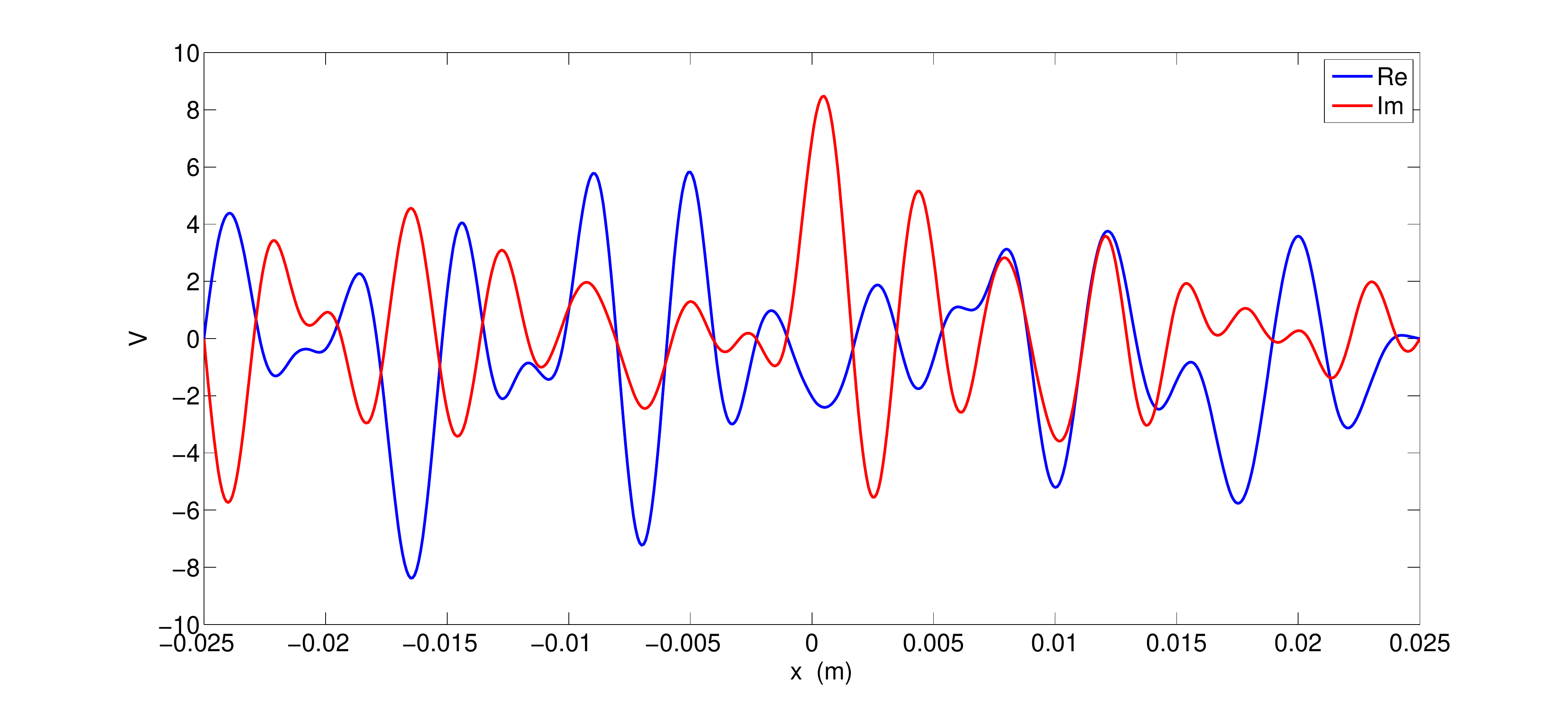}
   \caption{
  The sum to $p=9$ of $\hat E_{sp}(k,s,x)$  at $y=0$, at $5$m into the straight after the bend ($s=5.55~$m) for $kR=5\cdot10^{5}.$}
   \label{fig:esbend5mstraight}
\end{figure}

In Fig.\ref{fig:esbend5e5} we see something like the whispering gallery picture, in which the field is concentrated near the outer wall and nearly absent at the inner wall.
The concentration would be more pronounced if the bend were longer.  Following  the $\hat E_s$ field into the following straight section we find that this pattern disappears, and the field spreads over the whole width of the chamber. Fig.\ref{fig:esbend1mstraight} and  Fig.\ref{fig:esbend5mstraight} give views at $1$~m and $5$~m along the straight. Notice a signal of increasing complexity in that the initial $90^o$ phase shift between real and imaginary parts has disappeared by $5$~m.
\subsection{The $p$-dependent Shielding Cutoff at Low Frequencies}
The present scheme works best at large $k$. As we see in (\ref{sva}), the slowly varying amplitude approximation stands to fail at sufficiently small $k$. Evidently a low frequency cutoff is required, but experience showed that this cannot be independent  of $p$. We seek guidance on how to place the cutoff from the analytically soluble model of a complete circular torus with rectangular cross section\cite{warnock-morton}. The wave equation that defines that model is the Bessel equation with source,
\be
\frac{\ptl^2\fp}{\ptl x^2}+\frac{1}{x+R}\frac{\ptl\fp}{\ptl x}
+\bigg(\gamma_p^2-\big(\frac{kR}{x+R}\big)^2\bigg)\fp=\hat S_p \ ,  \label{bessel}
\ee
where definitions are the same as in (\ref{bendeq}) except that $kR=n$ is quantized to be an integer, so that solutions are periodic around the torus.
The corresponding homogeneous equation has solutions $J_n(\gamma_p(x+R))\ ,\ Y_n(\gamma_p(x+R))$ and resonances occur near frequencies where  $J_n(\gamma_p(x_++R))=0$ or $J_n\pr(\gamma_p(x_++R))=0$. Solutions at resonance are called whispering gallery modes and are concentrated near the outer wall and are very small at the inner wall.
Bessel functions have oscillatory behavior, allowing zeros at the outer wall, only when their argument is greater than their order. The {\it transition} from exponential to oscillatory behavior, where argument equals order, coincides with vanishing of the coefficient of $\fp$ in (\ref{bessel}). This is analogous to the change in solutions
of the harmonic equation $\ddot{u}+\omega^2u=0$ when $\omega^2$  changes sign. Thus a necessary condition for resonances is that
\be
\gamma_p^2 > \frac{(kR)^2}{(x_++R)^2}\ .    \label{trans}
\ee
Invoking the definitions and making an expansion for small $x_+/R$ we cast this in the form
 \be
 kR>\frac{\alpha_p(x_++R)}{\big(\beta^2(1+x_+/R)^2-1\big)^{1/2}}~\approx~ \pi p\frac{R}{h}\big(\frac{R}{2x_+}\big)^{1/2} \ ,\ (\beta=1)\ . \label{shcut}
 \ee
 This  ``shielding cutoff" for a toroidal or pillbox chamber was noted in \cite{warnock-morton}~(Eq.5.15 of SLAC-PUB-4562 or Eq.90 of the published paper).
 For $p=1$ and $2x_+=h$ it agrees with the better known cutoff for the parallel plate model of the vacuum chamber, derived in \cite{warnock-plates}.
A detailed analysis showed that there are no significant contributions to wake fields from frequencies below the resonance region \cite{warnock-ng}.

 In our single-pass system with moderate bend angle there are no sharp resonances, but there are broad peaks in the frequency spectrum of fields that apparently
 are vestiges of resonances, and also field patterns with some resemblance to whispering gallery modes. The resemblance to the full torus sharpens as the bend angle increases.
 Our example from LCLS-II has both small bend angle and large bending radius in comparison to most examples studied in the past, and is relatively remote in behavior
 from the full torus or ``steady state". We should therefore check to see if the cutoff (\ref{shcut}) looks reasonable by comparing field patterns below and above cutoff.
 \begin{figure}[htb]
   \centering
   \includegraphics*[width=\linewidth]{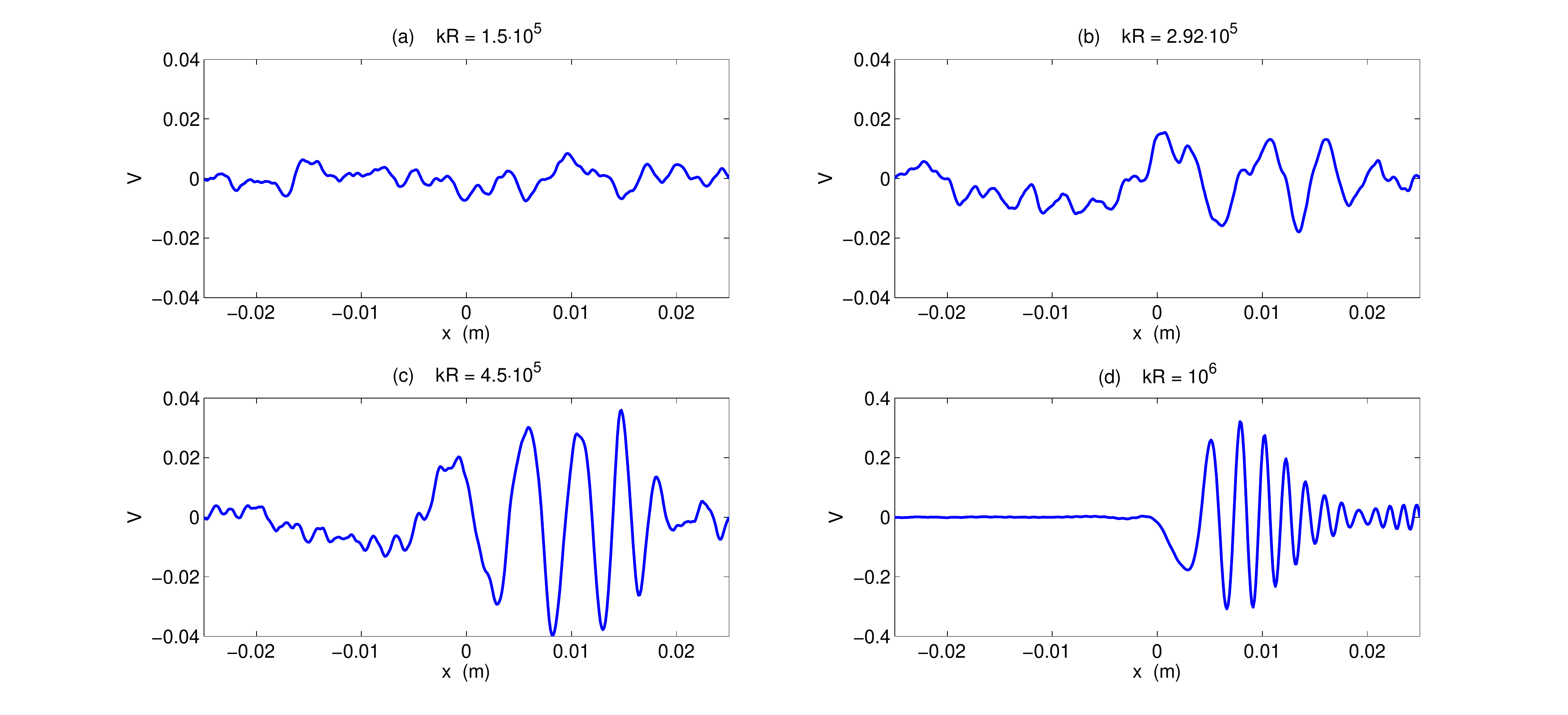}
   \caption{$\es(k,s,x)$ vs. $x$ for $s=0.55~$m at the end of the bend, $p=9$, comparing values of $kR$ around the shielding cutoff at $kR=2.92\cdot10^5$.  }
   \label{fig:shldcomp}
\end{figure}

In Fig.\ref{fig:shldcomp} we show results for $p=9$, usually the highest necessary vertical mode, plotting the longitudinal electric field as a function of $x$ as it looks at the end of the bend. Graph (b) is for $kR$ at the cutoff given by (\ref{shcut}).
The pattern far above cutoff in graph (d) can be considered a whispering gallery mode. Such a pattern appears to be emerging in graphs (b) and (c), but is absent in graph (a) below cutoff. Moreover, graph (a) is smaller, somewhat noisy, and shows poorer convergence as the mesh in $x$ is refined.  Notice the 10-fold larger scale in
 graph (d), which is for $k$ in the important spectral region for the wake field.   Similar results are found at smaller $p$.

 We conclude that a cutoff slightly less than (\ref{shcut}) will allow even incipient whispering gallery behavior, while excluding low frequency effects that will have negligible effect on wake fields and energy radiated. The excluded contributions are analogous to the small sub-resonant effects studied in \cite{warnock-ng}.  To check
 the effect of frequencies somewhat below the cutoff (\ref{shcut}), say down to the range of graph (a),  we insert a reduction factor $c_r$
 on the right hand side of (\ref{shcut}) and experiment with its value. We find that results for wake fields and energy radiated and absorbed are just the same  for $c_r=0.5$ and $c_r=1$. Thus it could be said that the vanishing of low frequencies effects is abrupt, just as in the full torus and parallel plate models.

 We incorporate the cutoff (\ref{shcut}) and find that our code has excellent convergence properties in all the parameters controlling discretization. This was not always the case with a $p$-independent cutoff.

\subsection{ Energy Radiated and Deposited in Resistive Walls}

\begin{figure}[htb]
   \centering
   \includegraphics*[width=.9\linewidth]{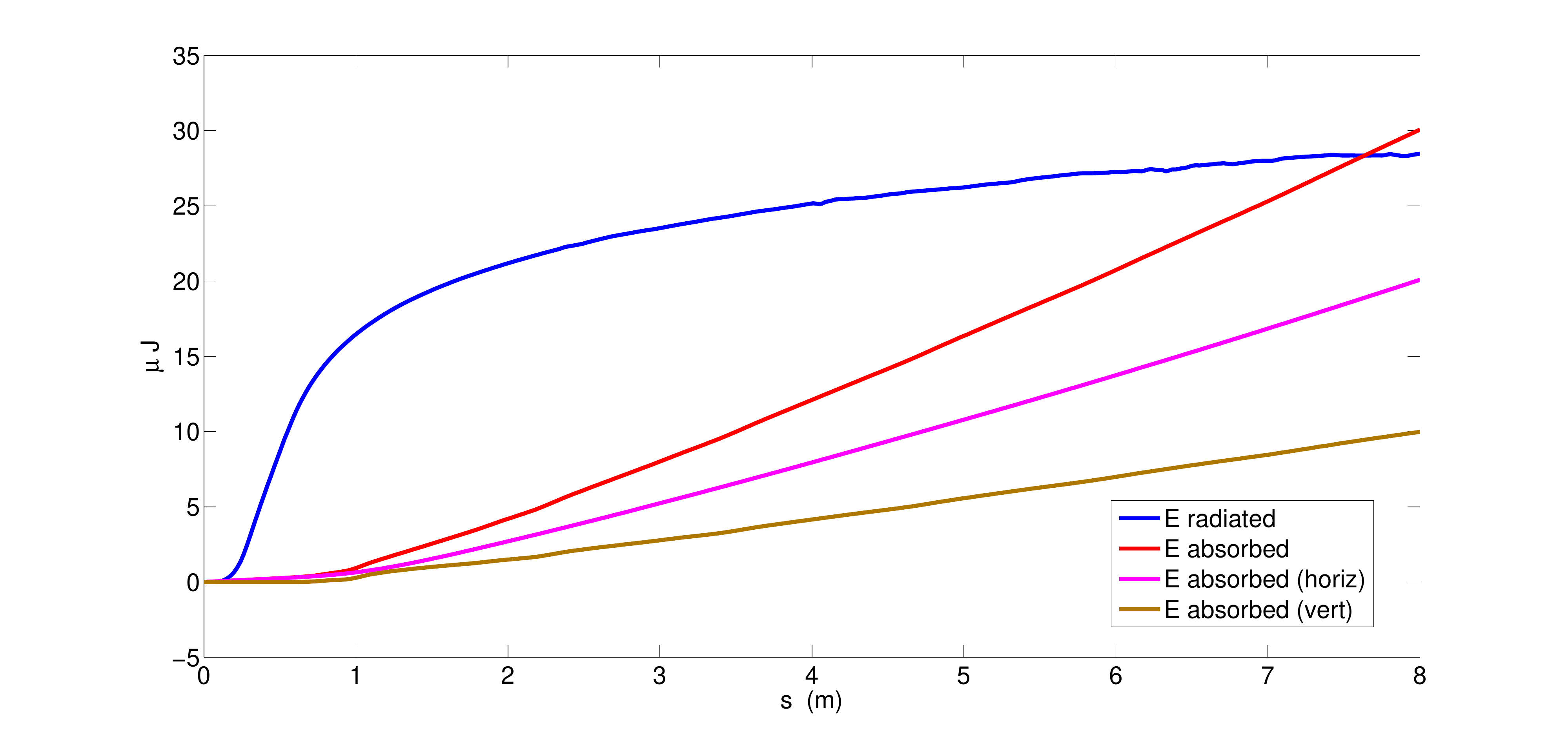}
   \caption{
  The total energy radiated in perfectly conducting model (blue); energy absorbed in perturbative model (red);
  energy absorbed in horizontal walls, perturbative model (magenta); energy absorbed in vertical walls, perturbative model (brown).
 The beginning of the bend is at $s=0$, where the fields have the steady-state values for an infinite straight waveguide.}
   \label{fig:eradabsorb}
\end{figure}

Having indicated  how the field components look in the frequency domain, we now turn to the integrals over all frequencies that give the energy radiated and absorbed in resistive walls. The blue curve in Fig.\ref{fig:eradabsorb} shows the total energy radiated up to position $s$ in the perfectly conducting model, as given by (\ref{eofs}).
The beginning of the bend is at $s=0$, where the fields are assumed to have the steady-state values for an infinite straight wave guide. The red curve
 Fig.\ref{fig:eradabsorb} shows the energy deposited in resistive walls up to position $s$ in the perturbative approximation.  This is obtained from the sum of the $s$-integrals of (\ref{hheat}) and (\ref{vheat}), for horizontal and vertical walls respectively. The magenta and brown curves give the separate contributions of horizontal and vertical walls.
\begin{figure}[htb]
   \centering
   \includegraphics*[width=\linewidth]{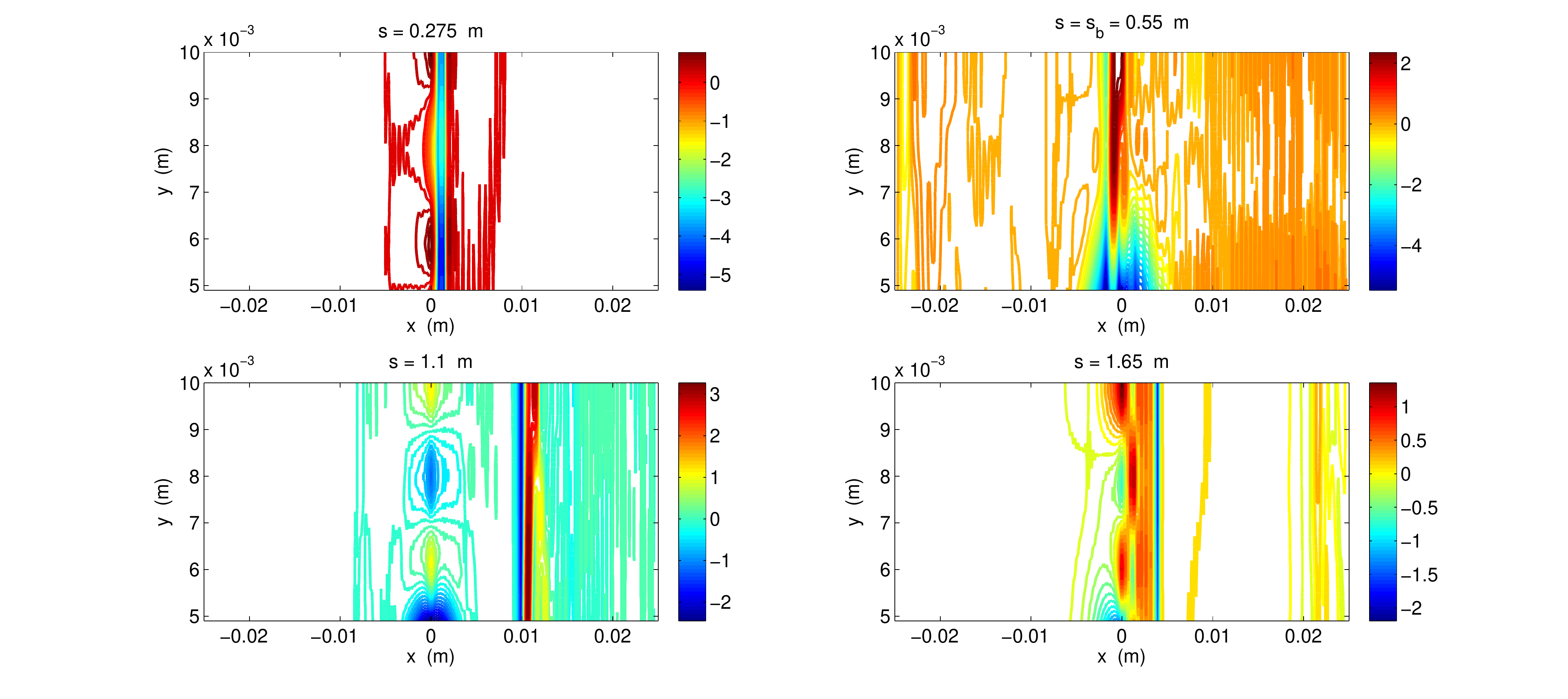}
   \caption{Contour plots of $H_x(s,x,y,t)$ in the $(x,y)$-plane at the instant the bunch is at location $s$. Plots in the upper $1/4$ of the chamber only, in
   arbitrary units. }
   \label{fig:contours}
\end{figure}
\begin{figure}[htb]
   \centering
   \includegraphics*[width=\linewidth]{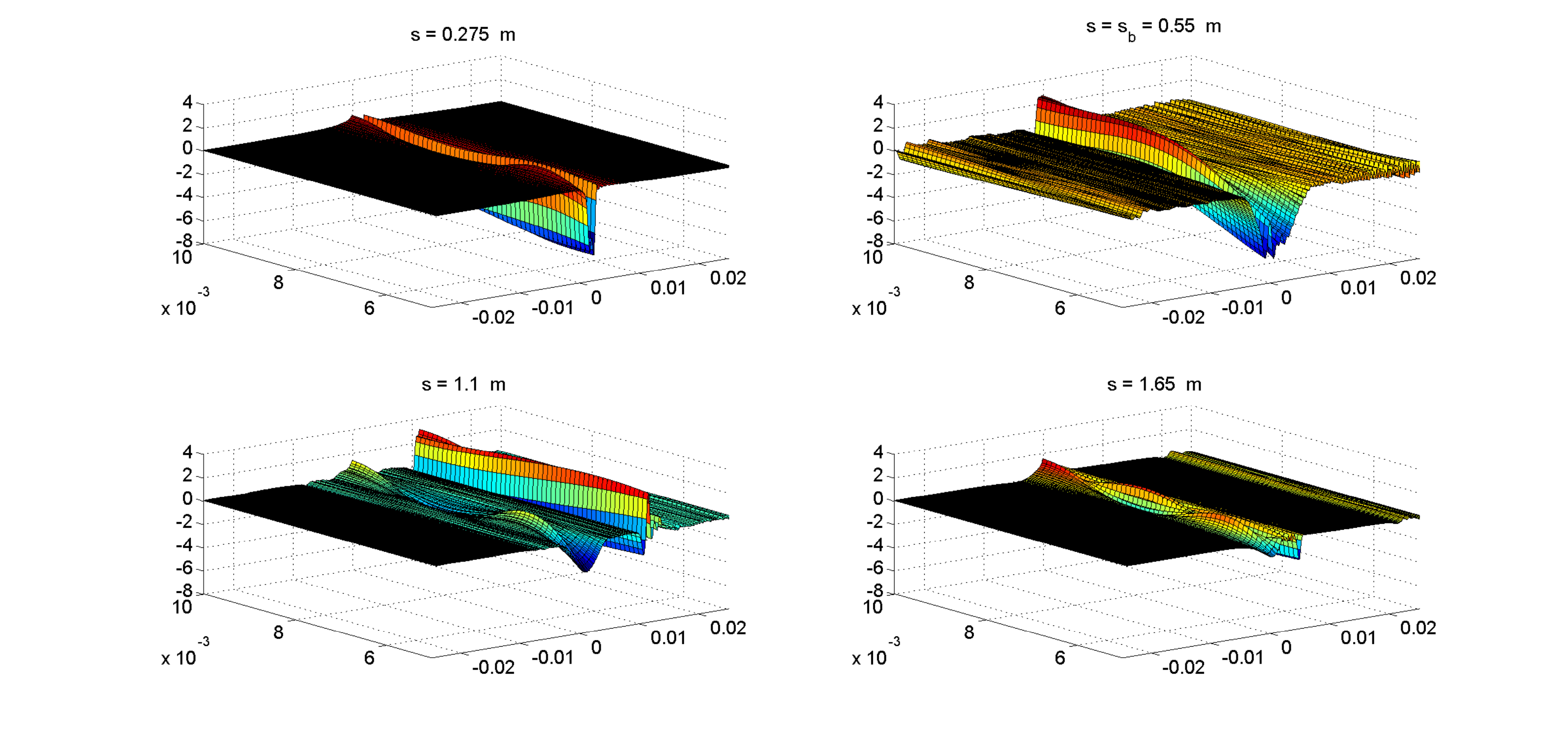}
   \caption{Surface plots of $H_x(s,x,y,t)$ as a function of $(x,y)$ at the instant the bunch is at location $s$. Plots in the upper $1/4$ of the chamber only,
   in arbitrary units. The $(x,y)$-domain is the same as that in Fig.\ref{fig:contours}.}
   \label{fig:surf}
\end{figure}
\begin{figure}[htb]
   \centering
   \includegraphics*[width=\linewidth]{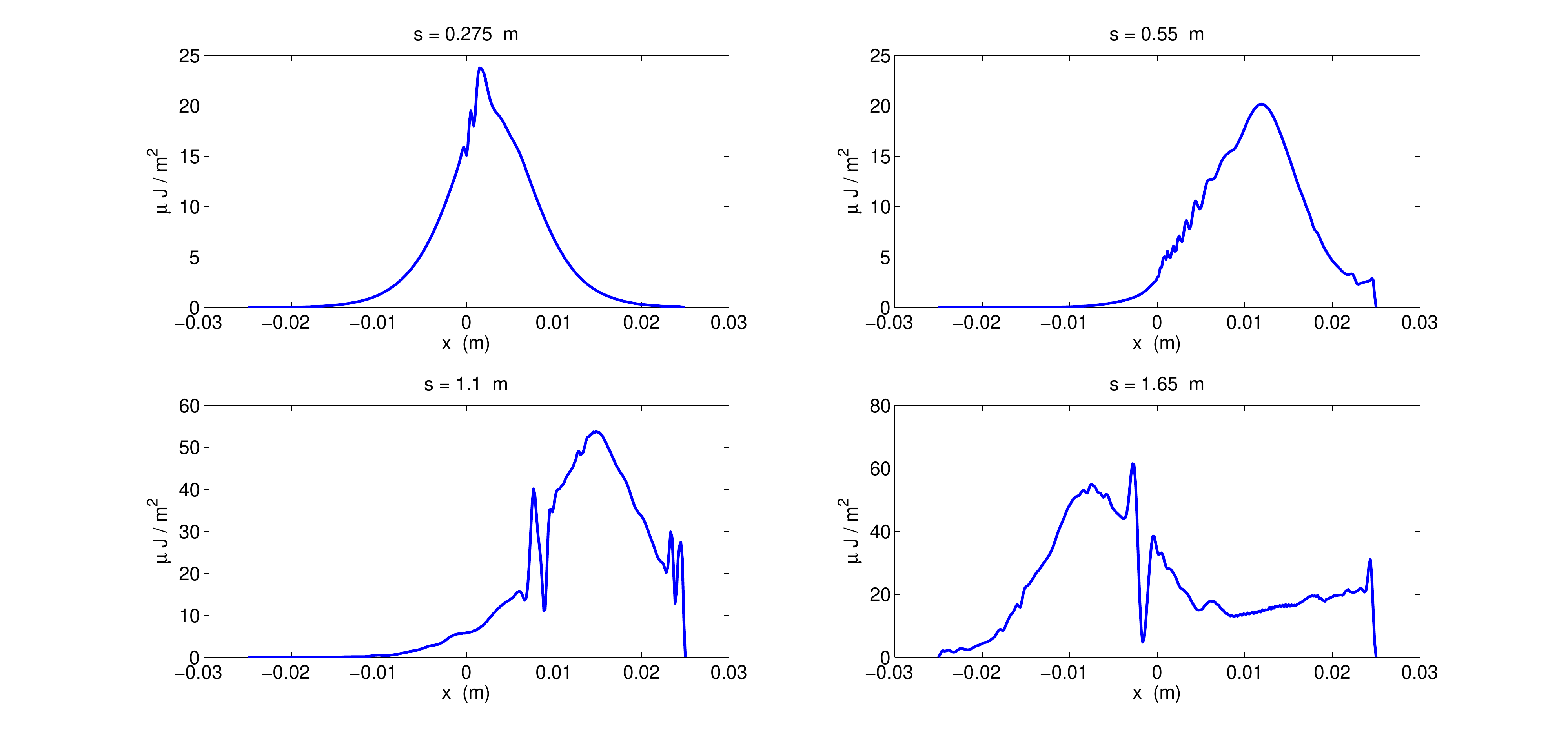}
   \caption{Time-integrated energy flow to the upper horizontal surface, per unit area, as a function of $x$ at various $s$.}
   \label{fig:flux_h}
\end{figure}

 The slope of the blue curve is nearly zero at $s=~$7.6 m, meaning that the decelerating field of CSR in the perfectly conducting model has nearly died out. This decay, a purely geometric effect, might be seen as a developing incoherence between different Fourier components of the longitudinal electric field at $x=0$. At large $s$ any one Fourier component is spread over all $x$, does not decay with $s$, and acquires a random looking relation between its real and imaginary parts.

  When the red curve crosses the blue curve at $s=s_c=7.6~$m all energy radiated at $s<s_c$ has been absorbed, to the accuracy of the perturbative model. In the present example, that happens to be almost all the energy that will be radiated.  With a lower wall conductivity it would be somewhat less than that, since the absorbed energy is proportional to $\sigma^{-1/2}$.

  Since a non-zero tangential $\bf H$ at the walls persists for $s>s_c$ the red curve continues to rise. This corresponds in exact physics to the fact that after all of the CSR is absorbed there is still a huge kinetic energy of the beam, some of which will be dissipated in the walls. Even though the perturbative model does not account self-consistently for decreasing energy of the beam, it is correct in predicting ongoing absorption in the walls. Asymptotically this should increase linearly with $s$, and indeed we find highly linear behavior of the red curve when the integration is extended by another $2~$m. The slope in that range is $5\ \mu {\rm J/m}$.

 Since the total CSR energy deposited from one bunch  is  $28\ \mu$J, and there are about $10^6$ bunches per second, the deposited power is about
 $28~$W. There is $5\ {\rm W/m}$ additional dissipated power for $s > s_c$.

 It is interesting to look at the field patterns in the region where a major part of the radiation occurs, say for $s<2~$m. In Fig.\ref{fig:contours} we show contour plots in $(x,y)$-space of
 $H_x(s,x,y,t)$ (in arbitrary units) at the instant the bunch passes position $s$.  We show this field component because its value on the horizontal walls  is the largest contributor to wall heating.
 We plot in just the upper quarter of the chamber, $g/2 < y < g$, since the range of values is too great to make a good plot in the full cross section. Corresponding surface plots are presented in Fig.\ref{fig:surf}.

  The flux of energy into the upper horizontal surface, integrated over time, is shown as a function of $x$ at various $s$ in Fig.\ref{fig:flux_h}. This function is given by (\ref{hheat}) without the integration on $x$.  The contribution of $\hat H_s$ to these plots is negligible.

\subsection{Longitudinal Wake Field and its Fourier Transform}
\begin{figure}[htb]
   \centering
   \includegraphics*[width=\linewidth,height=.35\linewidth]{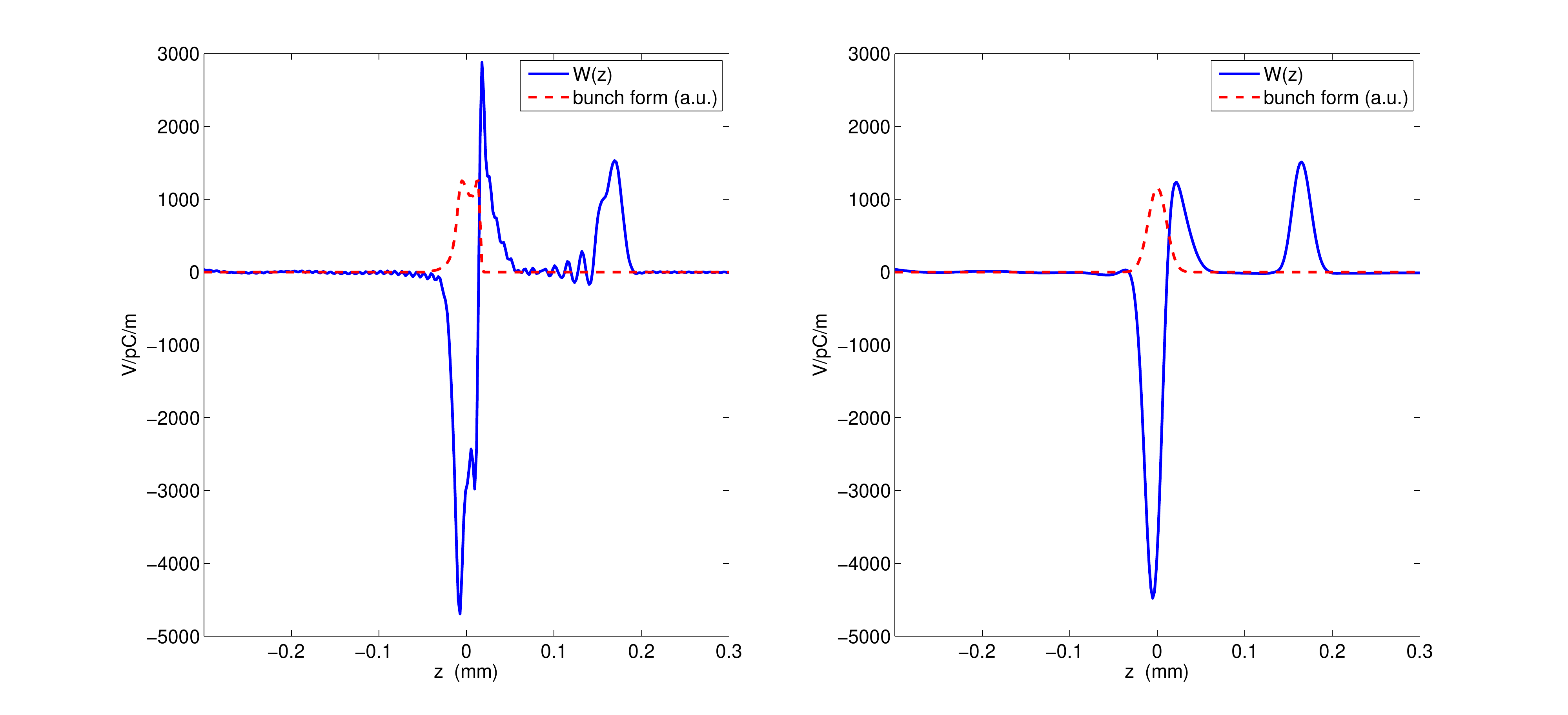}
   \caption{Wake field $W(z,s)$ as a function of $z=s-\beta ct$ at $s=0.55~$m (end of bend). Graph on left is for the simulated bunch form of Fig.\ref{fig:bunch}, that on right for a Gaussian bunch with the same $\sigma_z$.}
   \label{fig:wakecomp}
\end{figure}

Most earlier work on CSR has concentrated on fields at positions within the bunch. Usually only the longitudinal wake field is computed, but on occasion transverse forces have been studied as well \cite{bassi}. Following this tradition we plot in Fig.\ref{fig:wakecomp} the longitudinal wake field at the end of the bend, for  the simulated bunch form and a Gaussian with the same $\sigma_z$.
\begin{figure}[htb]
   \centering
   \includegraphics*[width=\linewidth,height=.35\linewidth]{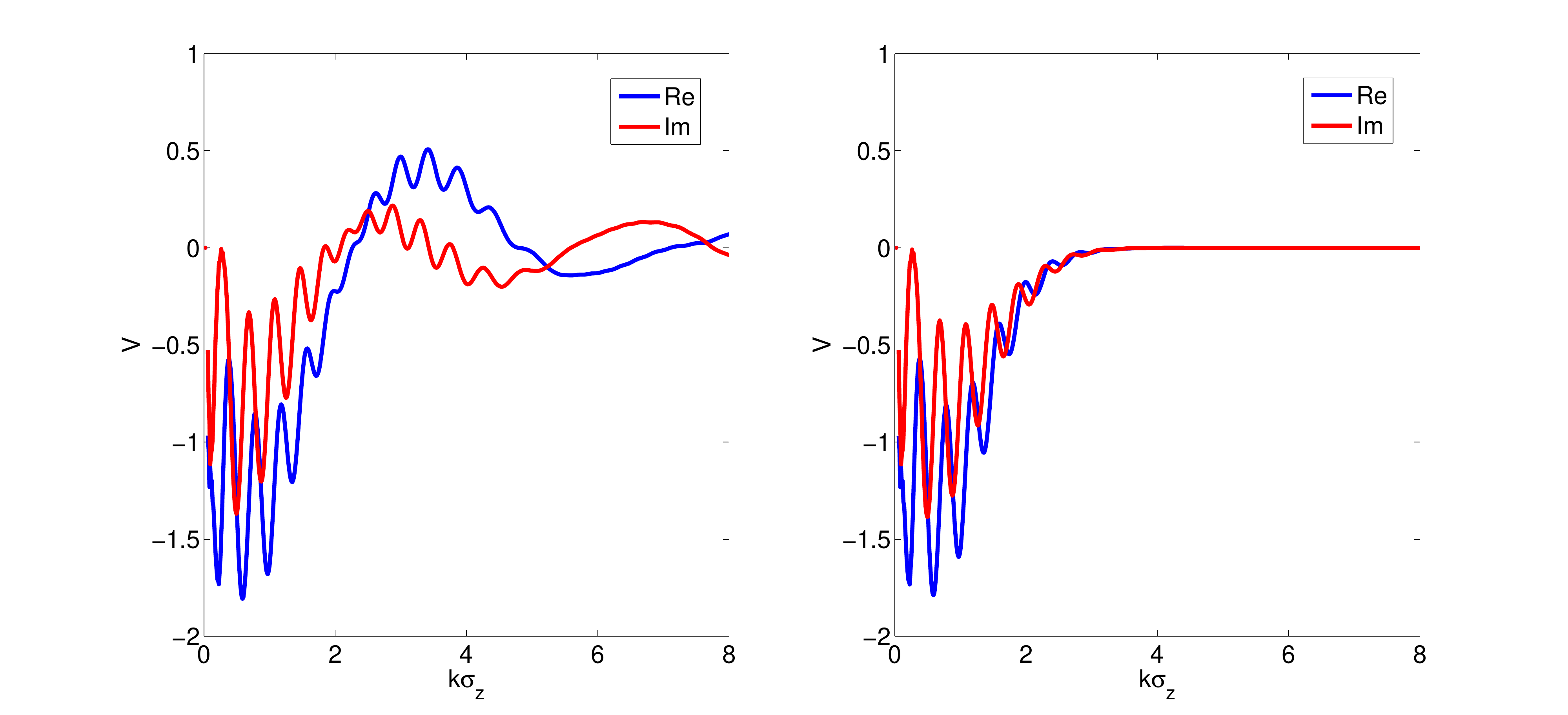}
  \caption{Fourier transforms of the two wake fields of Fig.\ref{fig:wakecomp}.}
   \label{fig:esk}
   \end{figure}
 \begin{figure}[htb]
   \centering
   \includegraphics*[width=\linewidth,height=.35\linewidth]{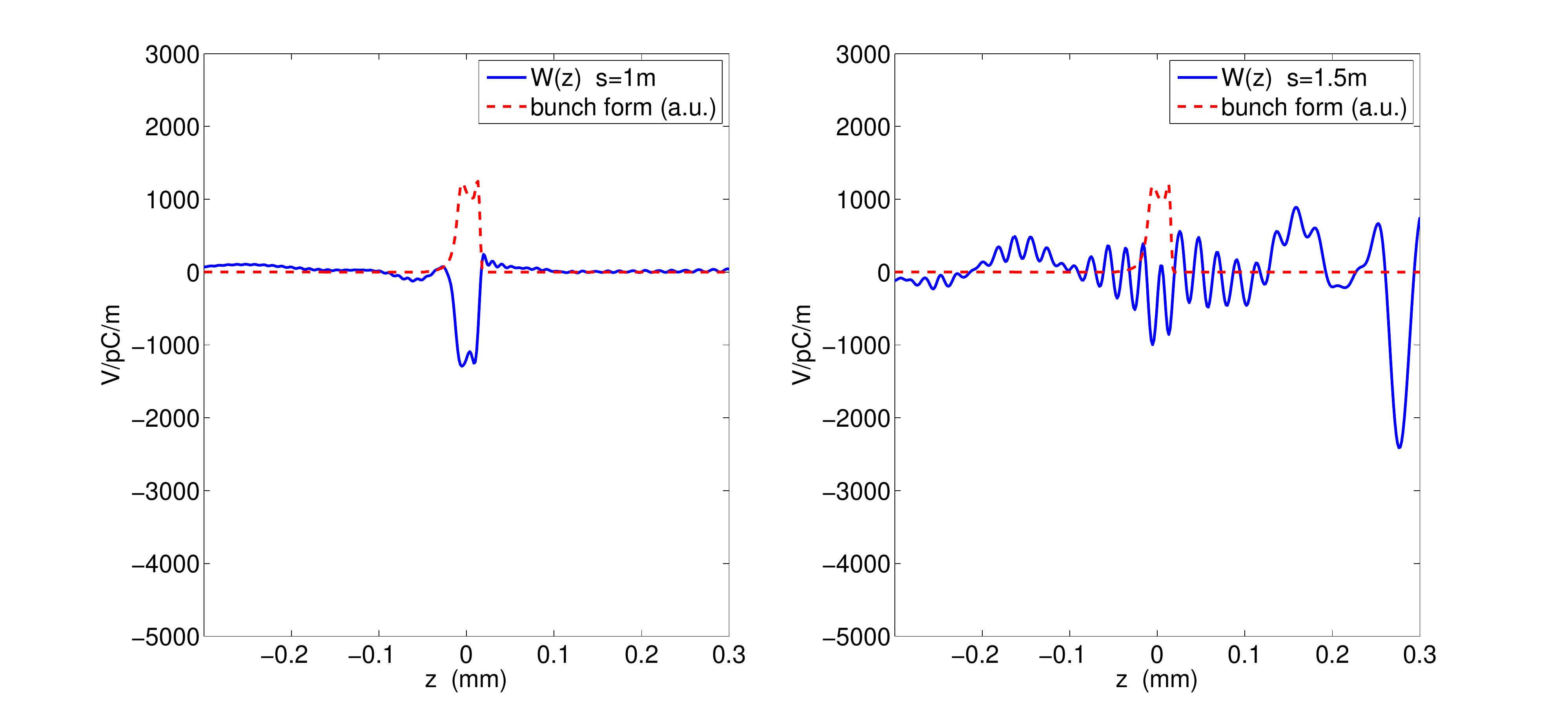}
   \caption{Wake field $W(z,s)$ as a function of $z=s-\beta ct$ at $s=1~$m (left) and at $s=1.5~$m (right).}
   \label{fig:wakedownstream}
\end{figure}
The dashed red curves represent the bunch form on an arbitrary scale. The head of the bunch is at positive $z=s-\beta ct$, and positive $W$ corresponds to energy gain.  We warn that this is just the electric field per unit charge at a fixed $s$ as a function of $z$, not the integral up to $s$ that is sometimes seen in the literature.

 \begin{figure}[htb]
   \centering
   \includegraphics*[width=.7\linewidth,height=.35\linewidth]{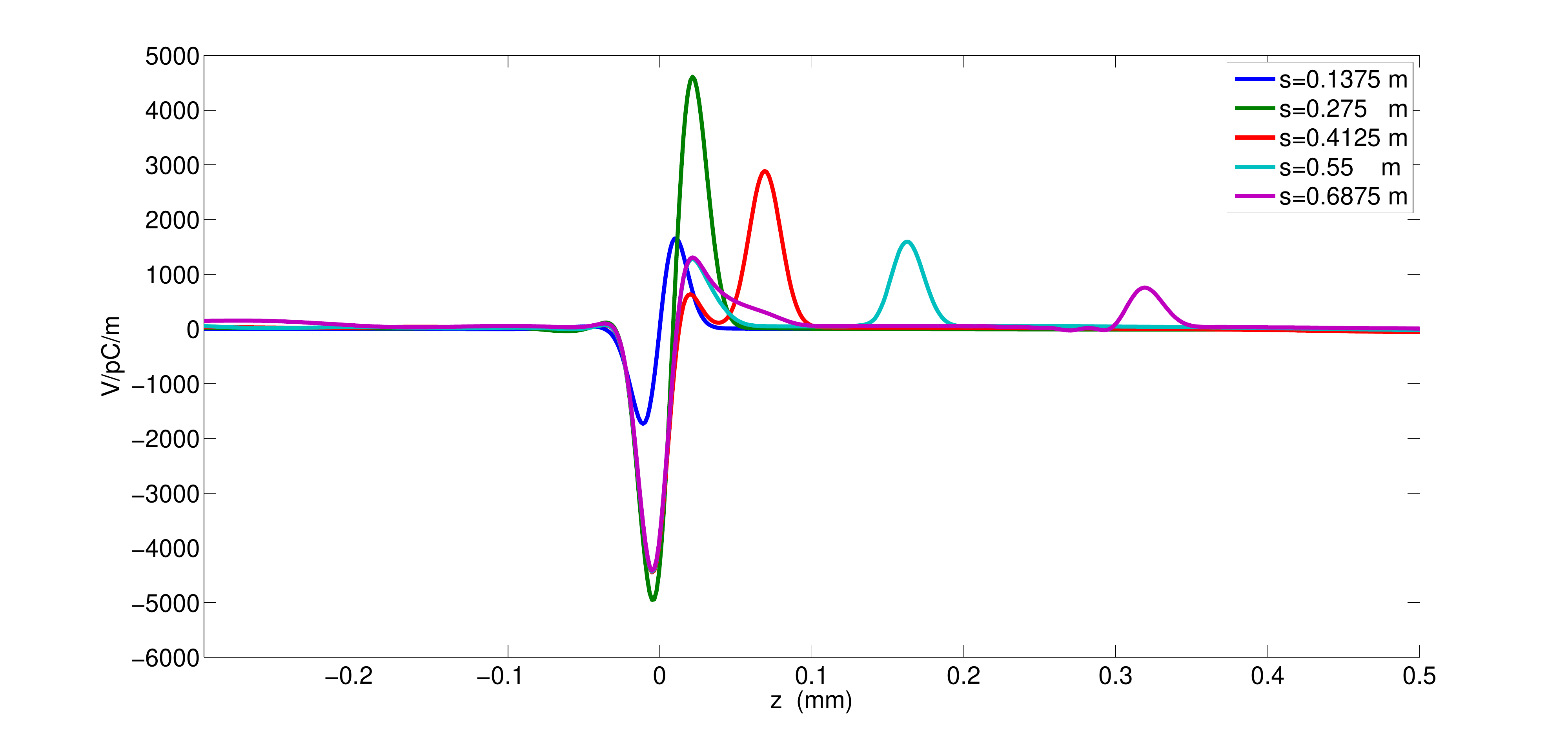}
   \caption{Evolution of the wake field $W(z,s)$ of a Gaussian bunch versus $s$ in a bend of length $s_b=0.825~$m.}
   \label{fig:forward_peak_evol}
\end{figure}

The corresponding Fourier transforms of $W(z)$ are displayed in Fig.\ref{fig:esk}.

The wakes at two values of $s$ beyond the bend are shown in Fig.\ref{fig:wakedownstream}.  With increasing $s$ there is a more and more complicated structure outside
 the support of the bunch, with decreasing energy loss within the bunch.

The peak several bunch lengths in front of the bunch in Fig.\ref{fig:wakecomp} is a parameter-dependent feature, which is not seen in several published plots for which the bend angle was larger \cite{stupakov-kotelnikov-2}, \cite{zhou}, \cite{zhou_etal}. A far-forward peak has turned up, however, in other publications; see Fig.3 and Fig.4 in \cite{agoh-yokoya} and Fig.17 in \cite{sagan}.  Furthermore, Stupakov and Emma did an analytic calculation of the wake field in free space \cite{stupakov-emma},
which showed such a peak evolving with position in the bend.

To show how a similar evolution works in our example for LCLS-II, we take a Gaussian bunch rather than the simulated bunch, to get a cleaner plot which is easier to understand.
We also extend the length of the bend by 50\% to $s_b=0.825~$m.  In Fig.\ref{fig:forward_peak_evol} we show the wake field at successive positions in the bend.
At $s=0.1375~$m the wake (blue curve) looks like the derivative of the bunch form. Subsequently the forward peak in this curve moves farther forward, and then splits into two.
The forward peak of the resulting pair moves ever farther from the bunch centroid, decreasing in height, and leaves the domain of the graph before $s=s_b$. Presumably, the
missing far-forward peak in certain publications is just out of range of the graphs.

\subsection{Code Validation and Timing}

We have applied two codes developed independently by the two authors. The results presented above are from the first code, which is written in Fortran and implements the
numerical methods and equations we have described. The second code is written in Matlab and is based on the same simplified wave equation and bunch description, but it makes a more sophisticated discretization in $x$, by the Discontinuous Galerkin (DG) Method
\cite{david-thesis}, \cite{hesthaven}. The DG scheme allowed first order smoothing of the source, as in (\ref{xi1def}), whereas second order smoothing was required for the finite difference method. Results of the two codes for the wake of the Gaussian in Fig.\ref{fig:wakecomp} agreed perfectly.  The DG method looks to be superior, but needs to be implemented in a faster programming language.

\begin{figure}[htb]
   \centering
   \includegraphics*[width=.6\linewidth, height= .3\linewidth]{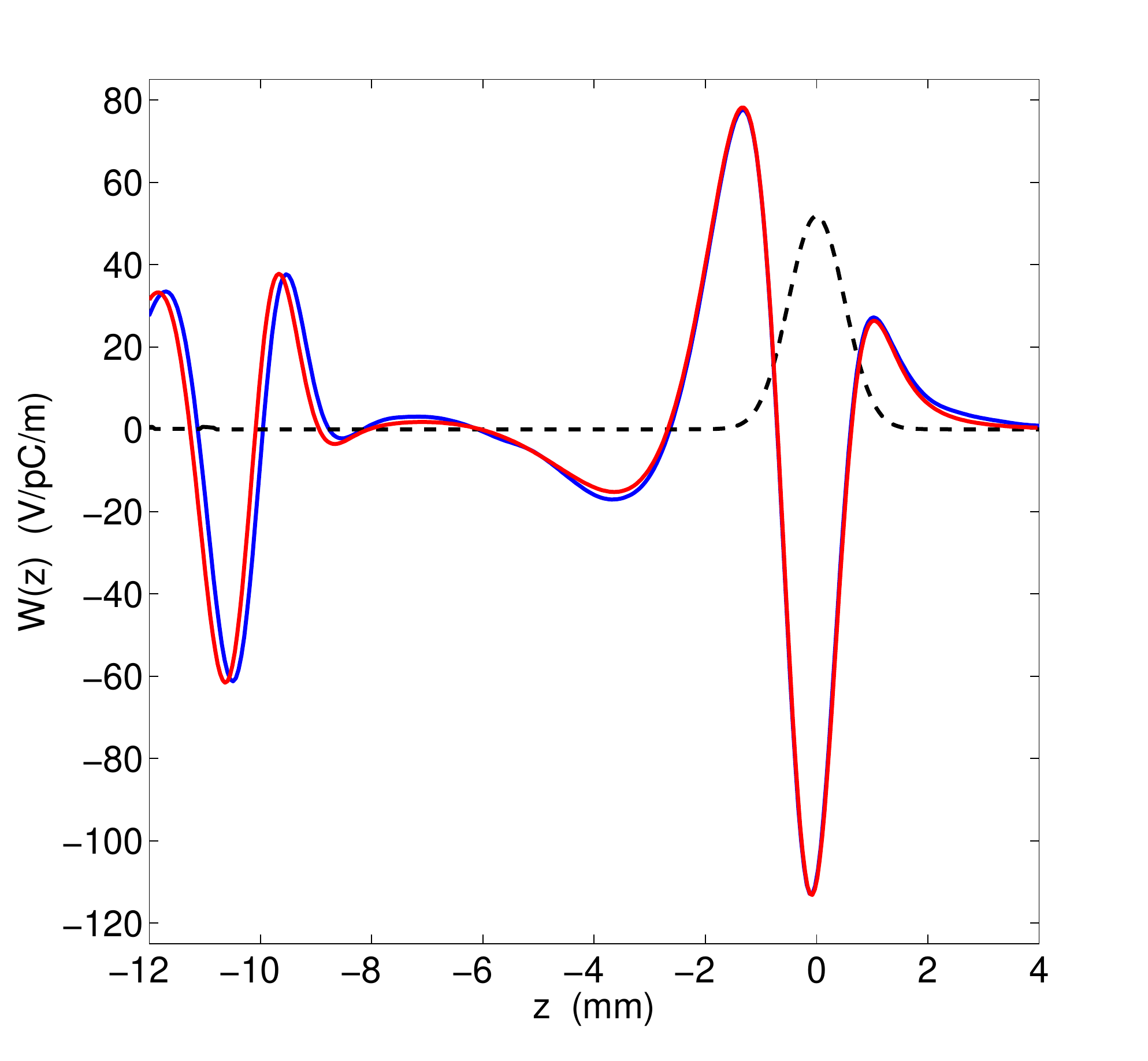}
  \caption{Wake field (blue) to compare with Fig.3b of \cite{stupakov-kotelnikov-2} (red).}
   \label{fig:skwake}
   \end{figure}

We have also made comparisons to results in the literature, for instance to the wake field in Fig.3b of \cite{stupakov-kotelnikov-2} for parameters $R=1~$m, $h=0.02~$m, $w=0.06~$m, with a Gaussian bunch having $\sigma_z=0.5~$mm.
This example was for the steady state case, computed by thorough mode expansions and requiring attention to poles on the real axis in frequency. To approximate the steady state we take a large bend angle of $\pi/2$ and find the wake at the end of the bend shown in the blue curve of Fig.\ref{fig:skwake}. The corresponding result  of \cite{stupakov-kotelnikov-2} is in red. Considering that the two calculations were done by vastly different methods, the close agreement is impressive.
We regard our calculation as simpler; for one thing, it does not require principal-value integrations around poles.

For the finite difference code the parameters controlling discretization are $\Delta s,\ \Delta x,\ p_{max},\ k_{max}, \ \Delta k$. By spot checks we have verified good convergence of various results in each of these parameters separately. The rate of convergence depends somewhat on the quantity computed. For the wake field at the end of the bend, in the LCLS-II example, a well-converged run could have parameters around the following:
\be
\Delta s = s_b/3\cdot 10^3\ ,\quad \Delta x = w/400\ , \quad p_{max}=9\ ,\quad k_{max}\sigma_z = 8\ ,\quad \Delta k = k_{max}/100\ ,   \label{discrt}
\ee
where $s_b=0.55~$m is the length of the bend. The integration step $\Delta s$ has more to do with stability of the $s$-integration than with accuracy. If it is small enough to ensure stability then making it even smaller  does not change results appreciably.  Examining the right hand side of (\ref{bendeq}) we see that an estimate for the Courant-Friedrichs-Lewy stability criterion is

\be
\Delta s <  2\alpha k(\Delta x)^2\ ,   \label{stabcrit}
\ee
where $\alpha < 1$ is to be determined empirically. We found that $\alpha=0.2$ worked in a few cases.
The value of $\Delta s$ in (\ref{discrt}) is for $\alpha=0.24$ and $k=k_{min}$ from the lowest shielding threshold. For simplicity our code takes all discretization parameters to be fixed, but it could be made more efficient by increasing $\Delta s$ with $k$, while keeping $\Delta x$ fixed.

With the control parameters of (\ref{discrt}) the computation of the wake field in Fig.\ref{fig:wakecomp} (left) took 4.2 minutes on a PC (Intel i7-4790, 3.6 GHz)
and included
evaluation of the wake field at 400 $z$-points at each of 400 different values of $s\le s_b$. (It also included 400 evaluations of the energies radiated and deposited in resistive walls, not needed for beam dynamics.) These results were achieved with a sub-optimal serial code. In parallel processing the $s$-integration could be done independently for each mode pair $(k,p)$, with a trivial calculation of initial data.

To secure convergence of the wake field within the bunch, the important region for beam dynamics, one can make a choice of discretization parameters more economical than (\ref{discrt}), say by decreasing $k_{max}$ and increasing $\Delta x$, the latter allowing a bigger $\Delta s$. In any case the code timing
 seems very promising for an application of our algorithm as a field solver for a self-consistent macro-particle simulation, or even a Vlasov calculation.

 At large $s$ out to $8~$m  we have to take a smaller $\Delta k$ than in (\ref{discrt}) in order to get a smooth curve of energy loss as in Fig.\ref{fig:eradabsorb}, say  $\Delta k=k_{max}/400$.  Otherwise we get a curve with about 3\% jitter at large $s$, meandering about the smooth curve.

\subsection{Check of the Slowly Varying Amplitude Approximation}
\begin{figure}[htb]
   \includegraphics*[width=\linewidth,height=.35\linewidth]{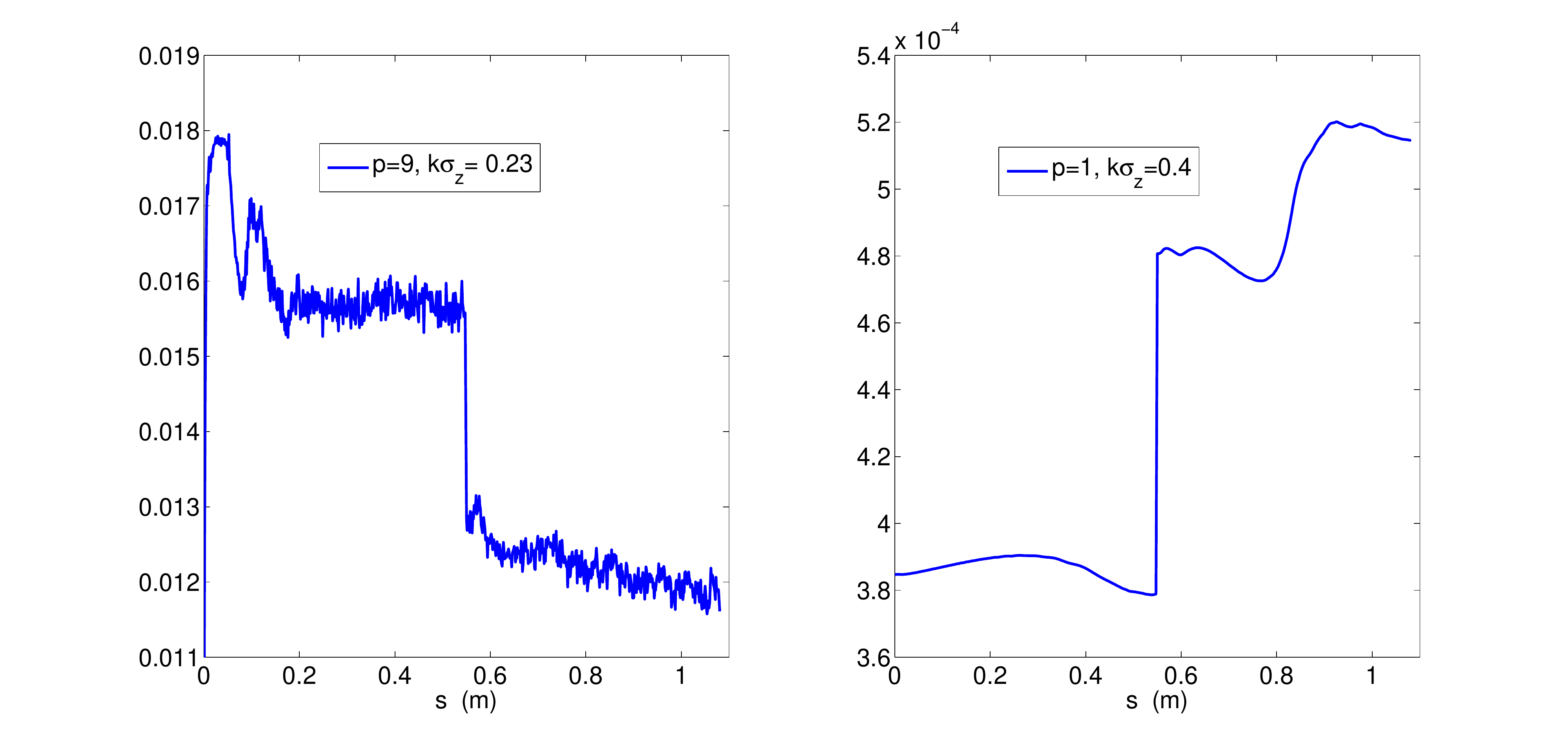}
  \caption{The ratio $r_E=\|\ptl^2\ey/\ptl s^2\|~/~2k\|\ptl\ey/\ptl s\|$ which should be small compared to 1 to justify the SVA approximation.}
   \label{fig:sva}
   \end{figure}

To verify the condition (\ref{sva}) for validity of the SVA approximation we take $\ptl\hat E_{yp}/\ptl s$ from the right hand side of (\ref{discreteb}) or
(\ref{discretes}), and approximate the second derivative as a divided difference. Then we plot the ratio $r_E(s)$ of the left hand side of (\ref{sva}) to the right hand side, and the similar ratio $r_H$ for the magnetic field. The ratios are largest at small $k$ and large $p$, so we put $p=9$ and $k$ at the shielding threshold (\ref{shcut}) for that $p$.  This ``worst case" for $r_E$ is shown in Fig.\ref{fig:sva} (left).  In a more important range of smaller $p$ and larger $k$ the values shown in Fig.\ref{fig:sva} (right) are  typical. The large step in values occurs at the bend-to-straight transition. Overall it appears that the SVA approximation is very well justified, for both $\ey$ and $\hy$, for the mild bend of the present example. For a large bend angle and small bend radius the justification is not so clear. For the case of Fig.\ref{fig:skwake} with $p=5$ and $k$ at the shielding threshold we find $r_E=0.13$ near the beginning of the bend.
\section{Conclusions and Outlook}
We have described an effective  scheme for fast numerical computation of CSR in a  vacuum chamber of rectangular cross section.
The central step is solving a simple system  of linear ordinary differential equations which describe evolution in $s$ of a slowly varying wave amplitude. The system is autonomous  within a bend, Eq.(\ref{discreteb}), or within a straight section, Eq.(\ref{discretes}). The source term $\tilde S_p$ is a continuous (or piecewise-continuous) function of $x$, obtained from the original line charge source by a change of dependent variable.

 We have applied our method to find the pattern
of energy deposited in resistive walls. To our knowledge, this effect of CSR was not previously studied.  We have treated the Poynting flux to the wall
just to lowest order in resistivity, which is good enough to establish that the effect is small.  An interesting next step would be a direct solution of the Maxwell equations
under the resistive wall boundary condition, a calculation required to find the rigorous resistive wall wake field.  This has been done analytically for the toroidal model \cite{warnock-morton},  and for special geometries with rectilinear beams \cite{bane}, \cite{bart},  but never for CSR in a vacuum chamber with successive bends and straights.

In this study we have learned three basic techniques of our chosen computational scheme: effective source smoothing by a change of dependent variable , the proper treatment of high $p$, and the proper low frequency cutoff. It remains to refine the numerical integration algorithms in several directions, and to take advantage of parallel processing.

A promising project is to apply this method as a field solver for self-consistent macro-particle simulations with a very short bunch, for instance for fields in a chicane including space charge as well as CSR. This should include a test of kernel smoothing \cite{silverman} as an elegant alternative to conventional particle-in-cell procedures for smoothing the charge/current density. Our method also offers opportunities to improve the study of CSR and beam dynamics in storage rings, especially to clarify the question of inter-bunch communication through long range wake fields from whispering gallery resonances \cite{bob-jack-marit}.

\section{Acknowledgments}
We thank James Ellison and Jack Bergstrom for a great deal of help and inspiration  over many years.
Paul Emma raised the question of resistive wall heating, and kindly provided a bunch simulation. Gennady Stupakov suggested the perturbative approach
to energy flux, gave us  good advice, and provided results of his code to compare with ours. The use of Frenet-Serret coordinates in this
study came out of conversations with Rui Li. Our work was supported  in part by U.S. Department of Energy contracts DE-AC03-76SF00515 and DE-FG02-99ER41104.

\appendix
\section{ Resistive Wall Boundary Condition}
We recall the derivation of the resistive wall boundary condition, adapting it to our particular context. We find the wave equation for the magnetic
field within the wall material, which is assumed to have  magnetic permeability $\mu$, electric permittivity $\epsilon$, and conductivity $\sigma$ all independent of position and frequency. The basic assumption is that the variation of the field within the wall is by far the strongest in the direction normal to the wall. This picture can be checked {\it a posteriori} by first assuming it to be true, then deducing the consequent normal variation. This variation, characterized by a small penetration depth (skin depth)
 can be compared with estimates of variation in the tangential directions.

Invoking Ohm's Law $\Jb=\sigma\Eb$, we have the curl equations within the wall as
\bea
&&\curl~\Hb=\sigma\Eb+\epsilon\frac{\ptl\Eb}{\ptl t}\ ,\label{hcurl}\\
&&\curl~\Eb=-\mu\frac{\ptl\Hb}{\ptl t}\ .\label{ecurl}
\eea
Next we take the Fourier transform with respect to $t$ to obtain

\bea
&&\curl~\tilde\Hb=(\sigma-i\omega\epsilon)\tilde\Eb\ ,\label{fthcurl}\\
&&\curl~\tilde\Eb=i\omega\mu\tilde\Hb ,\label{ftecurl}
\eea
where
\be
\tilde F(\omega,\rb)=\frac{1}{2\pi}\int dt~e^{i\omega t}F(\rb,t)\ ,\quad \rb=(s,x,y)\ .  \label{ftdef}
\ee
The term $-i\omega\epsilon$ from the displacement current is tiny in comparison to $\sigma$ at the highest frequencies we consider, and will be dropped henceforth.

We define a positive depth coordinate $\xi$, the distance from the beginning of the wall to an interior point of the wall medium, and a unit vector $\nb$ normal to the wall and directed from the wall toward the vacuum. At the horizontal walls $\xi=\pm(y-g)$, whereas at vertical walls $\xi=\pm(x-x_\pm)$. Then with the assumption of dominant normal variation the gradient is represented as $\nabla=-{\bf n}\ptl/\ptl\xi$, so that
\be
-\nb\times\frac{\ptl\tilde\Hb}{\ptl\xi}=\sigma\tilde\Eb\ ,\quad -\nb\times\frac{\ptl\tilde\Eb}{\ptl\xi}=i\omega\mu\tilde\Hb\ .\label{dxieqs}\\
\ee
 We can then eliminate $\tilde\Eb$ in (\ref{dxieqs}) by taking the curl of the first equation and substituting the second:
\be \nb\times\frac{\ptl}{\ptl\xi}\bigg(\nb\times \frac{\ptl}{\ptl\xi}\tilde\Hb\bigg)=\bigg(\nb\cdot\frac{\ptl^2\tilde\Hb}{\ptl\xi^2}\bigg)\nb-(\nb\cdot\nb)\frac{\ptl^2\tilde\Hb}{\ptl\xi^2}=
i\omega\mu\sigma\tilde\Hb\ .  \label{elime}
\ee
Since $\nabla\cdot\tilde\Hb=\ptl(\nb\cdot\tilde\Hb)/\ptl\xi=0$, we have
\be
\frac{\ptl^2\tilde\Hb}{\ptl\xi^2}+i\mu\sigma\omega\tilde\Hb=0\ . \label{heq}
\ee

The general solution of this harmonic equation with complex frequency is
\be
\tilde\Hb={\bf a}_+\exp(\xi/\Delta)+ {\bf a}_-\exp(-\xi/\Delta)\ ,\quad  \Delta^{-1}=(-i\mu\sigma\omega)^{1/2}\ .\label{homo}
\ee
The ${\bf a}_\pm$ depend only on coordinates other than $\xi$.
Since the solution must decay at large  $\xi>0$ we retain only the second term and choose the branch of the square root so that
${\rm Re}\Delta>0$, namely as
\be
\Delta^{-1}=e^{-i\pi/4}(\mu\sigma\omega)^{1/2}=(1-i)\big(\frac{\mu\sigma\omega}{2}\big)^{1/2}\ ,   \label{Deldef}
\ee
where the square root in (\ref{Deldef}) is positive at positive real $\omega$. We define this root in the complex $\omega$-plane with a cut
on the positive real axis. It then acquires a factor of $i$ in analytic continuation to negative $\omega$, so that $\Delta^{-1}$ has positive real
part at negative as well as positive $\omega$.
The conventional skin depth $d$ is defined by
\be
\Delta^{-1}=(1-i)/d\ , \quad d=\bigg(\frac{2}{\mu\sigma\omega}\bigg)^{1/2}\ , \label{deldef}
\ee
so that the field decays by a factor $1/e$ in a distance $d$.

By (\ref{homo}) we have $\ptl\tilde\Hb/\ptl \xi=-\tilde\Hb/\Delta$ which when substituted in the first equation of (\ref{dxieqs}) yields
\be
\tilde\Eb=(1-i)\bigg(\frac{\mu\omega}{2\sigma}\bigg)^{1/2}\nb\times\tilde\Hb\ .  \label{prebc}
\ee
 Taking the limit $\xi\rightarrow 0$ in (\ref{prebc}) we have the resistive wall boundary condition, since there must be
continuity with the fields in the vacuum.

 The Fourier transform
(\ref{ftdef}) with respect to time is related to the slowly varying amplitude $\hat F$  by the phase factor $\exp(-iks)/\beta c$, which cancels out in
(\ref{prebc}).
Thus with $\omega=\beta kc$ and the near-perfect approximation $\mu=\mu_0$, the boundary condition for the Fourier amplitudes used in this paper is
\be
\hat \Eb(k,s,x,y)=(1-i)\bigg(\frac{\beta Z_0k}{2\sigma}\bigg)^{1/2}\nb\times\hat\Hb(k,s,x,y)\ ,\label{finalbc}
\ee
at every point $(s,x,y)$ on the boundary, with the unit normal $\nb$ to the boundary  directed toward the vacuum.
The skin depth with $\mu=\mu_0$ is $d=(2/\beta Z_0 k\sigma)^{1/2}$.

\begin{figure}[htb]
   \includegraphics*[width=\linewidth,height=.35\linewidth]{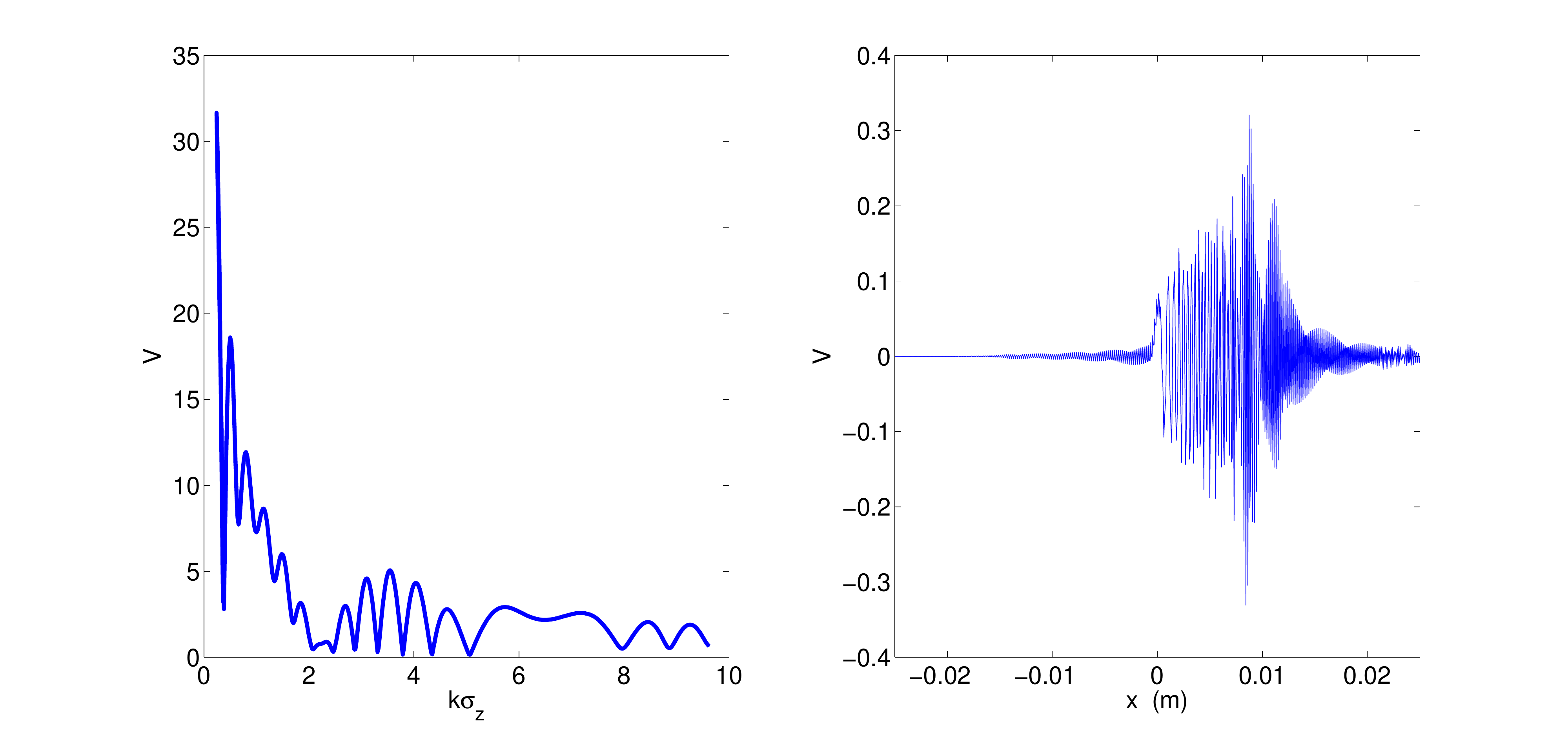}
  \caption{$Z_0|\hat H_x(k,s_b,0,g)|$ as a function of $k$ (left) and $Z_0\hat H_x(k,s_b,x,g)$ as a function of $x$ at $k\sigma_z=8$ (right).}
   \label{fig:hxspec}
   \end{figure}

To test the assumption of dominant normal variation, we assume that any tangential variation would not be faster within the wall than it is  at the surface. We  can
then estimate the scale of transverse variations at the surface from the perfectly conducting model, and compare it to the skin depth.
We first compare $d$ to the scale of variation in the $s$-direction, which should be about $1/k$. Since $d$ decreases with increasing $k$, an upper bound on
$d/k^{-1}$ will be its value at $k_{max}$, the largest relevant $k$. To decide on the latter we examine the Fourier spectrum of field components, for instance
 $|\hat H_x(k,s_b,0,g)|$ as a function of $k$ as plotted in Fig.\ref{fig:hxspec} (left), or the bunch spectrum in Fig.\ref{fig:lambk}. The most important range of the
 spectrum is for $k\sigma_z<3$, but there are substantial contributions out to $k\sigma_z=8$ or more.  Taking $k_{max}\sigma_z=8$ we find

\be
\frac{d}{k^{-1}}<\big[\frac{2k_{max}}{Z_0\sigma}\big]^{1/2} < 0.0083\ .  \label{svar}
\ee
For  $y$-variation the  corresponding ratio of interest is
\be
\frac{d}{\alpha_p^{-1}}< \alpha_p \big[\frac{2}{Z_0k_{min}(p)\sigma}\big]^{1/2} =\bigg[\frac{w}{R}\bigg]^{1/4}\bigg[\frac{2\alpha_p}{\pi Z_0\sigma}\bigg]^{1/2}=1.7\cdot10^{-5}p^{1/2}\ , \label{yvar}
\ee
where we use the shielding threshold (\ref{shcut}) for $k_{min(p)}$. Recall that $p_{max}=9$ in our calculations.
For corresponding estimates of variation in the $x$-direction, we refer to the numerical calculation of $\hat H_x(k,s_b,x)$ as a function of $x$. That has rapid oscillations when $k$ is large, as shown in Fig.\ref{fig:hxspec} (right) for $k\sigma_z=8$.  We define $\delta x(k)$ as $1/4$ of the period of the oscillations.  A rough fit shows that $\delta x(k)$ decreases more quickly than $d(k)$, at about the rate $k^{-1.125}$,
which means that $d(k)/\delta x(k)$ will have its maximum value at $k_{max}$. Reading off $\delta x$ from graphs for $k=k_{max}=8/\sigma_z$ we find
\be
\frac{d(k)}{\delta x(k)}<0.021\ .  \label{yvar}
\ee
This decreases to $0.014$ at $k\sigma_z=3$.
 To summarize, it appears that the scale of tangential field variations is at most about $2\%$ of the skin depth for the parameters of our example. This
 justifies the assumption of dominant normal variation, but not by the huge margin that might have been expected.

\end{document}